\documentclass[twocolumn]{autart}    

\usepackage{amsmath,amssymb,amsfonts}
\usepackage{setspace}
\usepackage{epstopdf}
\usepackage{indentfirst}
\usepackage{algorithmic}
\usepackage{graphicx}
\usepackage{algorithm}
\usepackage{svg}
\usepackage{caption}
\usepackage{textcomp}
\usepackage{float}    
\usepackage{natbib}
\usepackage{epsfig}
\usepackage{array}
\usepackage{mathrsfs}
\usepackage{color}  
\usepackage{enumitem}
\usepackage{hyperref}
\hypersetup{
	colorlinks=true,
	linkcolor=blue,
	filecolor=blue,
	urlcolor=blue,
	citecolor=blue,
}
\usepackage{longtable}
\begin{document}
	
	\begin{frontmatter}
		\title{Reinforcement Learning based Constrained Optimal Control: an Interpretable Reward Design \thanksref{footnoteinfo}}  
		
	 \thanks[footnoteinfo]{This work is supported by the National Natural Science Foundation of China under Grants 62233005 and U2441245. }

		\author[1]{Jingjie Ni},
		\author[1,2]{Fangfei Li},
		\author[2,3]{Xin Jin},
		\author[1]{Xianlun Peng},
		\author[2]{Yang Tang}

		\address[1]{School of Mathematics, East China University of Science and Technology, Shanghai 200237, China.}     
		\address[2]{Key Laboratory of Smart Manufacturing in Energy Chemical Process, Ministry of Education, East China University of Science and Technology, Shanghai 200237, China.}     
		\address[3]{Research Institute of Intelligent Complex Systems, Fudan University, Shanghai 200433, China.}

		
		\begin{abstract}                          
			This paper presents an interpretable reward design framework for reinforcement learning based constrained optimal control problems with state and terminal constraints. The problem is formalized within a standard partially observable Markov decision process framework. The reward function is constructed from four weighted components: a terminal constraint reward, a guidance reward, a penalty for state constraint violations, and a cost reduction incentive reward.
			A theoretically justified reward design is then presented, which establishes bounds on the weights of the components. This approach ensures that constraints are satisfied and objectives are optimized while mitigating numerical instability.
			Acknowledging the importance of prior knowledge in reward design, we sequentially solve two subproblems, using each solution to inform the reward design for the subsequent problem. Subsequently, we integrate reinforcement learning with curriculum learning, utilizing policies derived from simpler subproblems to assist in tackling more complex challenges, thereby facilitating convergence.
			The framework is evaluated against original and randomly weighted reward designs in a multi-agent particle environment. Experimental results demonstrate that the proposed approach significantly enhances satisfaction of terminal and state constraints and optimization of control cost.
		\end{abstract}
		
	\end{frontmatter}
	
	\setstretch{0.98}
	\setlength{\parskip}{1ex} 
	\setlength{\jot}{0pt} 
	\section{Introduction}
	The free-terminal-time optimal control problem, characterized by constraints on both terminal and state conditions, is highly pertinent in practical applications, as illustrated in Table \ref{tb:problem}. In addressing this, methods such as the calculus of variations \cite{variations2018}, Pontryagin's Maximum Principle \cite{Pontryagin2019}, dynamic programming \cite{Dynamic2013}, and reinforcement learning (RL) \cite{Reinforcement} are frequently employed. Among them, RL is highlighted owing to its model-free nature, promise to deal with the curse of dimensionality,  multi-agent collaboration capability without explicit policy communication, and efficient real-time computation \cite{bertsekas2019reinforcement}. Despite RL's success in complex decision-making, its application to state and terminal constraints with free terminal conditions is still developing \cite{gu2024review}. Next, we will concisely outline RL methods for handling state and terminal constraints.
	
	\begin{table*}[htb]
		\centering
		\caption{Practical applications of free terminal time optimal control with terminal and state constraints}
		\label{tb:problem}
		\fontsize{5.75pt}{6pt}\selectfont 
		\begin{tabular}{|c|c|c|c|}
			\hline
			\textbf{Problem} & \textbf{Terminal Constraints} & \textbf{State Constraints} & \textbf{Optimization Goals} \\
			\hline
			Motion planning \cite{MPcontrolcost, MPtime} & Given position & Collision-free & Fuel or terminal time  \\
			\hline
			Formation reconfiguration \cite{FR}& Relative positions & Physical limits and collision-free & Terminal time \\
			\hline
			Spacecraft rendezvous \cite{SR} & Rendezvous distance and safe speed & Collision-free & Fuel \\
			\hline
			Attitude control \cite{PL}& Given attitude and angular velocity & Physical limits of attitude  & Terminal time \\
			\hline
		\end{tabular}
	\end{table*}

	In the context of state constraints, safe RL  offers an effective approach to uphold safety constraints while optimizing cumulative rewards. For an extensive review of safe RL advancements, we direct readers to the comprehensive survey \cite{gu2024review}. Among various safe RL techniques \cite{han2021reinforcement,  tessler2018reward}, the penalty-based methods receive great attention due to their nature as first-order methods and their simplicity in implementation. 
	Existing penalty-based methods \cite{2022saute, IPO2020, IPO2024, P2BPO2024} have demonstrated empirical effectiveness; however, they either lack rigorous theoretical justifications or may compromise the consistency between the primal and penalty-based problems. The penalized proximal policy optimization method \cite{P3O2022} uniquely ensures theoretical equivalence between the original and penalty-based optimization frameworks, provided the penalty factor exceeds the maximum Lagrange multiplier. However, it does not specify the appropriate range for the penalty factor, which is a critical issue. An excessively large penalty factor can lead to numerical instability and reduced performance \cite{2022saute, P3O2022}, while a factor that is not sufficiently large may fail to maintain equivalence. 
	To address numerical instability from large penalty factors, some studies \cite{yan2022relative} have employed curriculum learning (CL) \cite{cl2009}, starting with relaxed state constraints and gradually increasing their strictness. This approach mitigates instability but does not solve the core issue of establishing explicit lower bounds for penalty weights. Within the optimization domain, proposals for explicit lower bounds on penalty weights in inequality-constrained contexts \cite{cluster2021} have been put forth. However, it presumes the differentiability and convexity of the objective function and prior knowledge that is challenging to obtain in model-free settings.
	In summary, identifying a computationally feasible lower bound for the penalty factor remains a significant challenge.

	In addressing terminal constraints, researchers often utilize reward shaping \cite{MPcontrolcost, safaoui2024safe, MPtime, FR, SR, PL}. Typically, a positive reward is awarded to the agent for successfully meeting terminal constraints. However, this reward is sparse, failing to effectively incentivize agents to transition towards terminal states before reaching them. To facilitate effective exploration, researchers introduce guidance rewards (dense rewards) to encourage agents to approach terminal states. For example, in terms of motion planning \cite{MPcontrolcost, MPtime, safaoui2024safe}, the closer an agent gets to the target, the higher the reward it receives.
	Nevertheless, the existing research has not yet identified a precise method for calibrating the magnitude of these two rewards. Similar to the penalty factor, a low reward for terminal constraints may lead agents to neglect these constraints, while an excessively high reward could risk numerical instability and cause agents to prioritize these rewards over other objectives. A minimal guidance reward may be eclipsed by other rewards, rendering the learning of terminal constraints and other objectives with equal priority inefficient \cite{huang2022reward}. Conversely, an excessively high guidance reward might mislead agents, as its maximization may not be equated to terminal state achievement \cite{ng1999policy,booth2023perils}.
	
	In summary, addressing state and terminal constraints individually poses challenges in setting the penalty factor, guidance rewards, and positive rewards for terminal constraints. We frame these challenges within the broader context of reward configuration issues. The lack of a standardized design approach leads to several concerns \cite{hayes2022practical,booth2023perils}: 1. Inefficient policy learning:  As previously discussed, misaligned reward weightings can fail to effectively convey optimization goals and constraints to agents, resulting in suboptimal policies despite extensive training. 2. Costly reward calibration: The lack of theoretical guidance necessitates a trial-and-error approach for reward tuning and policy validation by engineers, which is intellectually demanding. 3. Lack of interpretability: Even when effective reward designs are identified for specific problems, the rationale behind the effectiveness of certain configurations over others remains unclear. This situation underscores an urgent need for our research to develop a reward design methodology for constrained optimal control scenarios.
	
	Taken together, this paper tackles the free-terminal-time optimal control problem with terminal constraints and state constraints, by introducing a theoretically sound and practically viable reward design approach. Our contributions are threefold:
	\begin{enumerate}[itemsep=0pt,topsep=0pt,parsep=0pt, leftmargin=*]
		\item  Building on \cite{P3O2022, MPtime}, we explicitly define the weight ranges for each reward component, including the lower bound for the penalty factor. We theoretically demonstrate that the optimal policy derived from our reward design is indeed the optimal policy for the original constrained control problem.
		\item  Compared to \cite{P3O2022, cluster2021}, the reward parameters in our approach can be readily obtained within the RL framework. Specifically, the necessary reward parameters are derived by solving a series of subproblems, with each solution informing the design of rewards for the subsequent problem. Subsequently, curriculum learning (CL) is employed to apply policies from simpler subproblems to aid in solving more complex ones, thereby facilitating convergence.
		\item  We tested our reward design framework in an open-source multi-agent particle environment \cite{lowe2017multi, hu2024distributed}. Compared to the original reward settings \cite{lowe2017multi, hu2024distributed} and those that did not effectively apply our theorems, ours achieved higher rates of state and terminal constraint satisfaction and demonstrated superior performance across optimization metrics.
	\end{enumerate}
	
	The structure of this paper is as follows: Section \ref{sec:pll} introduces the Multi-Agent RL approach and formulates the constrained optimal control problem within the Multi-Agent RL framework. Section \ref{sec:tg} details a theoretically grounded reward design. Section \ref{sec:reward} offers practical guidelines for reward design. Section \ref{sec:sim} provides empirical evidence supporting our proposed method. Finally, Section \ref{sec:con} is the conclusion.
	
	$\mathbf{Notations:}$ We use the following notations throughout this paper. $\mathbb{R}^{+} $, $\mathbb{R}$, $\mathbb{R}^{m}$ denote the sets of non-negative real numbers, real numbers, and $m$-dimensional vectors, respectively. $\mathbb{E}[\cdot] $ denotes the expected value operator. For \( x \in \mathbb{R}^m \), the notation $\|x\|_2$ signifies the 2-norm of \( x \).
	
	\section{Preliminary and Problem Formulation}\label{sec:pll}
	This section introduces the Multi-Agent RL approach and formulates the constrained optimal control problem within the Multi-Agent RL framework.
	
	\subsection{Preliminary}
	We present a Multi-Agent RL approach within a Partially Observable Markov Decision Process (POMDP) framework, formalized as $\langle n, \mathcal{S}, \mathcal{O}, \mathcal{A}, \mathcal{P}, \mathcal{R}, \gamma \rangle$. Here, \( n \) denotes the number of agents, and $\mathcal{S} = \{s^t, t \in \mathbb{Z}^+\}$ denotes the state space. The local observation for agent $i$ at time $t$ is denoted by $o_i^t = \mathcal{O}(s^t, i)$. The shared action space is $\mathcal{A} = \mathcal{A}_1 \times \mathcal{A}_2 \times \ldots \times \mathcal{A}_N$, with the individual action space for agent $i$ being $\mathcal{A}_i = \{a^t_i, t \in \mathbb{Z}^+\}$. The transition probability from state $s^t$ to $s^{t+1}$ given the joint action $a^t = (a^t_1, \ldots, a^t_N)$ is denoted by $\mathcal{P}(s^{t+1} \mid s^t, a^t)$. The shared reward function is $\mathcal{R}(s^t, a^t)$, and the discount factor is $\gamma$. The policy for agent $i$ is defined as $\pi_i: o_i^t \rightarrow a^t_i, \forall t$. We define the agents’ joint policy as $\pi= \pi_1\times\pi_2\times ...\times \pi_n$.
	In this paper, we consider a fully cooperative setting where all agents share the same reward function, aiming to maximize the expected return 
	\begin{equation}
		J(s^0, \pi) = \mathbb{E}[\sum_{t=0}^{T_f} \gamma^t \mathcal{R}(s^t, a^t)], 
		\label{eqj}
	\end{equation}
	where $T_f$ is the terminal step. The optimal joint policy can be obtained by integrating it with the general Multi-Agent RL algorithm \cite{mappo2022}. 
	
	\subsection{Optimal Constrained Control Problem}
	We consider a free-terminal-time optimal control problem for a discrete-time system of \( n \) agents, subject to terminal and state constraints. 
	Let \( x_i(t) \in \mathbb{R}^p \) and \( u_i(t) \in \mathbb{R}^q \) represent the control state and control input of agent \( i \), respectively. Define \( x(t) = [x_1(t)^{\top},x_2(t)^{\top},...,x_n(t)^{\top}]^{\top} \in \mathbb{R}^{np} \) be the joint control state and \(u(t) =  [u_1(t)^{\top},u_2(t)^{\top},...,u_n(t)^{\top}]^{\top} \in \mathbb{R}^{nq} \) be the joint control input. Agent $i$ can observe the control states of its $m$ neighbors $\overline x_{i}(t) = [x_{i1}(t)^{\top}, x_{i2}(t)^{\top}, \ldots, x_{im}(t)^{\top}]^{\top}$, where $x_{ij}(t)$ is the control state of the $i^\text{th}$ neighbor of agent $i$. Our goal is to determine a policy $\pi_i: [x_i(t)^{\top}, \overline x_{i}(t)^{\top}, t]^{\top} \rightarrow u_i(t)$ for each agent $i = 1, 2, \ldots, n$ at any time step $t$ that satisfies the following conditions:
	\begin{subequations}
		\label{eq:P2}
		\begin{align}
			&\min_{\pi_1, \pi_2,\dots,\pi_n}  \sum_{t=1}^{t_f}c\big(x(t),u(t)\big), \label{optC} \\
			\text{s.t. } 
			&\label{initial2} x(0) \in \Phi,\\
			&\label{motion2} x(t+1) = f\big(x(t),u(t)\big) , \  t = 0,..., t_f, \\
			&\label{goal2}  x(t_f)\in F, \\
			&\label{tmax2} t_f < t_{\max}, \\
			&\label{safe2} x(t)\in C, \  t = 0,..., t_f, \\	
			&\label{a2} u(t)\in U, \  t = 0,..., t_f.
		\end{align}
	\end{subequations}
	Our optimization objective is presented in (\ref{optC}),where $c:\mathbb{R}^{np} \times \mathbb{R}^{nq}\rightarrow \mathbb{R}^{+}$ represents the cost function. The initial joint control state is constrained to the set $\Phi \subseteq \mathbb{R}^{np}$, defined by (\ref{initial2}). 
	The discrete-time system dynamics are governed by the deterministic transition function $f: \mathbb{R}^{np} \times \mathbb{R}^{nq} \rightarrow \mathbb{R}^{np}$, as described in (\ref{motion2}).
	The terminal state must remain within the set $F \subseteq \mathbb{R}^{np}$, as indicated by the terminal constraint in (\ref{goal2}). 
	The control horizon is finite, limited to $t_{\max}$, as specified in (\ref{tmax2}). The state must remain within the set $C \subseteq \mathbb{R}^{np}$, as indicated by the state constraint in (\ref{safe2}). 
	The control input constraint is given in (\ref{a2}), where $U \subseteq \mathbb{R}^{nq}$ represents the feasible set of control inputs.
	The problem (\ref{eq:P2}) is prevalent in practice as shown in Table \ref{tb:problem}. The cost function in (\ref{optC}) can be tailored for diverse objectives \cite{lewis2012optimal}. For minimum-time, set
	\begin{equation}
		c\big(x(t),u(t)\big) = 1, \label{eqct}
	\end{equation} 
	which makes (\ref{optC}) equivalent to minimizing \( \sum_{t=1}^{t_f} 1 = t_f \). For minimum-fuel problems, use
	\begin{equation}
		c\big(x(t),u(t)\big) = \|u(t)\|_2. \label{eqcu} 
	\end{equation} 
	

	\subsection{POMDP Formulation}
	Now, we formalize the problem (\ref{eq:P2}) within the POMDP framework. First, we define $\langle\mathcal{S}, \mathcal{O}, \mathcal{A}, \mathcal{P}\rangle$. The details of $\langle\mathcal{R}, \gamma\rangle$ will be presented in subsequent sections. The state $s^t$ is defined as the concatenation of the joint control state and the current time step, denoted as $[x(t)^{\top},t]^{\top}$. 
	The local observation for agent $i$ at time $t$ is denoted by $o_i^t = [x_i(t)^{\top},\overline x_{i}(t)^{\top}, t]{\top}$.
	The action for agent $i$ is defined as $a^t_i = u_i(t)$. The transition probability $\mathcal{P}(s^{t+1} \mid s^t, a^t)$ is governed by the deterministic function $f$, as specified in (\ref{motion2}). 
	
	Next, we outline our approach to incorporating constraints (\ref{initial2})-(\ref{a2}) within the POMDP framework. First, the initial state constraint (\ref{initial2}) is ensured by specifying the initial state at the start of each episode.
	The system dynamics (\ref{motion2}) can be naturally incorporated into the RL framework via the transition probability function $\mathcal{P}$. 
	To handle the control input constraint (\ref{a2}), we consider several scenarios. If the control input is continuous and bounded, i.e., \( U = \left\{u(t) \big| |u_{ij}(t)| \leq \overline{u}_{ij}(t), i=1,\ldots,n, j=1,\ldots,q\right\} \), with \( |u_{ij}(t)| \) being the absolute value of the \( j \)-th element of \( u_i \) and \( \overline{u}_{ij}(t) \) representing its maximum allowable value, we can use an activation function like \texttt{tanh} in the output layer of the policy network to ensure compliance. For a discrete control set \( U = \{u^1, u^2, \ldots, u^k\} \), algorithms designed for discrete actions can be applied \cite{mappo2022}. For more complex forms, techniques such as action masking may be employed \cite{10812765}. 
	Finally, the terminal constraint (\ref{goal2}), the finite-time constraint (\ref{tmax2}), and the state constraint (\ref{safe2}) are guaranteed by defining an appropriate terminal step $T_f$ and $\langle\mathcal{R}, \gamma\rangle$. Specifically, for any time step, the terminal step $T_f$ is set to $t$ if one of the following conditions is met: (1). The terminal constraint in (\ref{goal2}) is satisfied; (2). The state constraint in (\ref{safe2}) is violated; (3). The current time step satisfies $t = t_{\max} - 1$, indicating that the next time step will violate (\ref{tmax2}). Next, we introduce the components $\langle\mathcal{R}, \gamma\rangle$.
	
	\section{Reward Design with Theoretical Guarantees}\label{sec:tg}
	
	For problem (\ref{eq:P2}), the reward typically consists of four components \cite{SR, MPtime, FR, PL}:
	\begin{equation}
		\begin{aligned}
			\mathcal{R}(s^t, a^t) &= \alpha\mathcal{R}^{a}(s^t, a^t) + \beta\mathcal{R}^{g}(s^t, a^t) + \lambda\mathcal{R}^{p}(s^t, a^t) \\
			&+ \mu\mathcal{R}^{c}(s^t, a^t).
		\end{aligned}
		\label{eq:reward2}
	\end{equation}
	The reward for meeting the terminal constraint is:
	\begin{equation}
		\mathcal{R}^{a}(s^t, a^t) = \left\{
		\begin{array}{ll}
			1, & \text{if\ } (\ref{goal2})  \text{\ is\ satisfied}, \\
			0, & \text{otherwise.}
		\end{array}\right.
		\label{eq:arreq}
	\end{equation}
	Here, we simply use 1 or 0 as the reward for whether equation (\ref{goal2}) is satisfied, a method that is general without loss of specificity, given that there is a weight $\alpha>0$ associated with $\mathcal{R}^{a}(s^t, a^t)$ in (\ref{eq:reward2}). To overcome sparse rewards, the guidance reward $\mathcal{R}^{g}(s^t, a^t)$ steers the agent towards the terminal state:
	\begin{equation}
		\mathcal{R}^{g}(s^t, a^t) =  l(s^t, a^t),
		\label{eq:naveq}
	\end{equation}
	where $l: \mathcal{S} \times \mathcal{A} \rightarrow \mathbb{R}$ is a guidance function, with $l(s^t, a^t)$ increasing as the state approaches the terminal state. $\beta\geq0$ is the weight of $\mathcal{R}^{g}(s^t, a^t)$. $\mathcal{R}^{p}(s^t, a^t)$ is the penalty for state constraint violation:
	\begin{equation}
		\mathcal{R}^{p}(s^t, a^t) = \left\{
		\begin{array}{ll}
			0, & \text{if\ } (\ref{safe2})  \text{\ is\ satisfied}, \\
			-1, & \text{otherwise.}
		\end{array}\right.
		\label{eq:safeeq}
	\end{equation}
	The constant $\lambda\geq$ 0 represents the weight of the penalty, with its magnitude reflecting the importance of state constraint satisfaction.
	The reward function $\mathcal{R}^{c}(s^t, a^t)$, aimed at minimizing the cumulative cost for agents, is defined as follows:
	\begin{equation}
		\mathcal{R}^{c}(s^t, a^t) =  c(s^t, a^t),
		\label{eq:rewardc}
	\end{equation}
	with $\mu \leq 0$ being the weight coefficient of $\mathcal{R}^{c}(s^t, a^t)$. Since the cost functions are equally weighted at each time step, in (\ref{eqj}), we set
	\begin{equation}
		\gamma = 1.
		\label{eq:gamma}
	\end{equation}

	To convey the goals of the proposed problem (\ref{eq:P2}) to the agents, it is crucial to define a sensible range for the reward parameters $\alpha$, $\beta$, $\mu$, and $\lambda$. To achieve this, we first establish Assumption 1 and then introduce Theorem 1, which provides sufficient conditions to ensure the consistency of the optimal joint policy that maximizes (\ref{eqj}) and the one for the proposed problem (\ref{eq:P2}).

	\noindent $\mathbf{Assumption\ 1}$.  There exists a positive constant $\rho \in \mathbb{R}^{+}$ such that for all state-action pairs $(s^t, a^t)$, the inequality $|l(s^t, a^t)| < \rho$ holds, where $l: \mathcal{S} \times \mathcal{A} \rightarrow \mathbb{R}$ is the guidance function.
	
	Prior to presenting Theorem 1, we define
	\begin{equation}
		\begin{aligned}
			\tau = \max_{s^{0},s^{t_f}}\mathbb{E}_{\tilde\pi}&\bigg[\sum_{t=1}^{T_f}c(s^t, a^t) \Big| s^{0} \text{\ satisfies\  (\ref{initial2})},\\
			&s^{t_f} \text{\ satisfies\  (\ref{goal2})} \bigg].
		\end{aligned}
		\label{eqtua}
	\end{equation}
	Here, the cumulative control cost upper bound $\tau$ is calculated for all potential initial states $s^{0}$ and terminal states $s^{t_f}$ under the joint policy $\tilde{\pi}$, which guarantees constraint satisfaction. Specifically, the constraint-satisfying joint policy $\tilde{\pi}$ ensures that all agents can satisfy the constraints (\ref{initial2})-(\ref{a2}), irrespective of the specific optimal objective (\ref{optC}).
	
	\noindent \textbf{Theorem 1}. For a noiseless kinematic system (\ref{motion2}), if $\beta$, $\mu$, and $\lambda$ satisfy the following conditions:
	\begin{equation}
		\lambda > \alpha,\quad \beta  = 0,\quad \mu  > - \frac{\alpha}{\tau},
		\label{eqpenbeta1Pphi}
	\end{equation}
	then the optimal joint policy derived from the reward (\ref{eq:reward2}) coincides with the optimal joint policy of the proposed problem  (\ref{eq:P2}).
	
	\noindent$\mathbf{Proof}$. See Appendix \ref{sec:T2}.
	$\hfill\blacksquare$

	Theorem 1 outlines the reward structure for an optimal policy in problem (\ref{eq:P2}), which requires the guidance reward weight $\beta$ to be zero and the identification of the parameter $\tau$. However, this setup is impractical due to the sparse reward issue caused by the lack of guidance rewards and the challenge of estimating $\tau$ without prior knowledge. To overcome these issues, we introduce a practical guide for reward design in the subsequent section.

	\section{Reward Design with Practical Guarantees}\label{sec:reward}
	In this section, we present a practical method for setting the guidance reward weight $\beta = 0$ and estimating $\tau$. Initially, we suggest addressing simpler problems to gather prerequisite knowledge for estimating the parameter $\tau$. Then, leveraging CL, we effectively sequence these simpler problems and skillfully utilize the policies derived from solving them to set $\beta = 0$ without affecting convergence.
	

	\subsection{Reward Parameter Acquisition Process by Solving Two Subproblems}
	This section describes our approach to acquiring the necessary knowledge for estimating the parameter $\tau$. To achieve this, we need to solve two subproblems: 1) the proposed problem (\ref{eq:P2}), which aims to minimize the time cost function (\ref{eqct}), known as the \textit{minimum-time problem}, and 2) the \textit{minimum-time problem} without state constraints (\ref{safe2}). We will then discuss the necessity of solving these subproblems and the methods used to address them.
	
	First, we concentrate on the \textit{minimum-time problem}. Upon resolution, we obtain the constraint-satisfying joint policy $\tilde{\pi}$. Using $\tilde{\pi}$, we guide agents to explore the environment and thereby determine $\tau$ (\ref{eqtua}). We now outline the solution to the \textit{minimum-time problem}.
	
	The reward and discount factor pair $\langle\mathcal{R}_m, \gamma_m\rangle$ for the \textit{minimum-time problem}, which, instead of aligning with (\ref{eq:gamma}) and (\ref{eq:reward2}), adopts a straightforward and ingenious form that facilitates the establishment of theoretical guarantees. Specifically, we set the weight $\mu$ of $\mathcal{R}^{c}(s^t, a^t)$ to zero, thereby simplifying equation (\ref{eq:reward2}) to:
	\begin{equation}
		\mathcal{R}_m(s^t, a^t) = \alpha\mathcal{R}^{a}(s^t, a^t) + \beta\mathcal{R}^{g}(s^t, a^t) + \lambda\mathcal{R}^{p}(s^t, a^t).
		\label{eq:rewardt}
	\end{equation}
	To minimize the terminal time $t_f$, we use a discount factor $\gamma_m < 1$ in (\ref{eqj}), encouraging agents to satisfy the terminal constraint promptly, as the discounted cumulative reward $\sum_{t=0}^{t_f} \gamma_m^t \mathcal{R}^{a}(s^t, a^t)$ increases with a decreasing $t_f$.
	
	For the \textit{minimum-time problem}, establishing appropriate parameter ranges for $\alpha$, $\beta$, and $\lambda$ is crucial. Thus, we introduce Theorem 2, providing conditions ensuring the consistency of the optimal joint policy that maximizes the cumulative of (\ref{eq:rewardt}) and the one for the \textit{minimum-time problem}. First, we define 
	\begin{equation}
		t_c = \min_{s^{0},s^{t_f}}\mathbb{E}_{\overline\pi}\bigg[t_f \Big| s^{0} \text{\ satisfies\  (\ref{initial2})}, s^{t_f} \text{\ satisfies\  (\ref{goal2})} \bigg].
		\label{eqtc}
	\end{equation}
	Here, \( \overline\pi \) is the optimal joint policy for the \textit{minimum-time problem} without considering the state constraint~(\ref{safe2}).
	
	\noindent $\mathbf{Theorem\ 2}$. For noiseless system model (\ref{motion2}), if $\beta$ and $\lambda$ satisfy the following conditions
	\begin{equation}
		\lambda>\alpha\gamma_m^{t_{c}-t_{\max}},\quad \beta<\frac{\alpha\gamma_m^{t_{\max}}(1 - \gamma_m)^2}{2\rho(1 - \gamma_m^{t_{\max}})}, 
		\label{eqpenbeta1}
	\end{equation}
	then the optimal joint policy derived from the reward design (\ref{eq:rewardt}) aligns with the optimal joint policy for the \textit{minimum-time problem}, specifically (\ref{eq:P2}) with (\ref{eqct}).
	
	\noindent$\mathbf{Proof}$. See Appendix \ref{sec:T1}.
	$\hfill\blacksquare$
	
	Theorem 2 establishes that the reward scheme results in an optimal joint policy $\tilde{\pi}$ for the \textit{minimum-time problem}, enabling the estimation of $\tau$. However, the value of $t_c$ within this reward scheme (\ref{eqpenbeta1}) remains undetermined. To ascertain $t_c$, akin to $\tau$, we must first solve for the joint policy $\overline{\pi}$ for the \textit{minimum-time problem} without state constraints (\ref{safe2}). This problem employs the same $\langle\mathcal{R}_m, \gamma_m\rangle$ framework as the \textit{minimum-time problem}, with the sole alteration being the setting of $\lambda = 0$ to account for the absence of state constraints. The specific reward parameter configurations are outlined in Corollary 1, with the proof methodologically analogous to Theorem 2.
	
	\noindent \(\mathbf{Corollary\ 1}\). For noiseless system model (\ref{motion2}), if \(\beta\) and \(\lambda\) meet the following requirements
	\begin{equation}
		\lambda = 0,\quad \beta < \frac{\alpha\gamma_m^{t_{\max}}(1 - \gamma_m)^2}{2\rho(1 - \gamma_m^{t_{\max}})}, 
		\label{eqpenbeta2}
	\end{equation}
	then the optimal joint policy derived from the reward design (\ref{eq:rewardt}) aligns with the optimal joint policy for the \textit{minimum-time problem} objectives without considering state constraints (\ref{safe2}).
	
	\noindent\(\mathbf{Proof}\). Refer to Appendix \(\ref{sec:T1}\).
	\(\hfill\blacksquare\)
	
	Based on Corollary 1, we can obtain the joint policy $\overline{\pi}$ for the \textit{minimum-time problem} without state constraints (\ref{safe2}). Subsequently, by directing all agents to explore the environment using $\overline{\pi}$, we can determine $t_c$ (\ref{eqtc}).
	
	Based on the analysis, computing $\tau$ involves three steps:
	1. Solve the \textit{minimum-time problem} without considering state constraint (\ref{safe2}) using Corollary 1 to obtain parameter $t_c$.
	2. Solve the \textit{minimum-time problem} using Theorem 2 to obtain parameter $\tau$.
	3. Solve the proposed problem (\ref{eq:P2}) using Theorem 1. The three-step process for obtaining reward parameters may seem complex. Next, we present the CL method, which minimizes the impact of the complex process on convergence speed and ensures that setting the guidance reward weight $\beta$ to zero does not hinder convergence.

	\subsection{A Practical Guide for Reward Design using Curriculum Learning}
	This section outlines the integration of CL~\cite{cl2009} with RL. CL sequences the training into progressively challenging stages, each building on the policies learned in prior stages. This method enables the agent to develop policies in less complex environments before advancing to more complex ones, thereby boosting learning efficiency. The addition of CL in the complex reward parameter acquisition process offers two advantages: 1) the application of policy reduces the impact of the complex process on convergence speed, and 2) it ensures that setting the guidance reward weight $\beta$ to zero does not hinder the process. Next, we present the curriculum design and its advantages in detail.
	
	\begin{figure}[H]
		\vspace{-5pt} 
		\centering
		\includegraphics[width=0.474\textwidth]{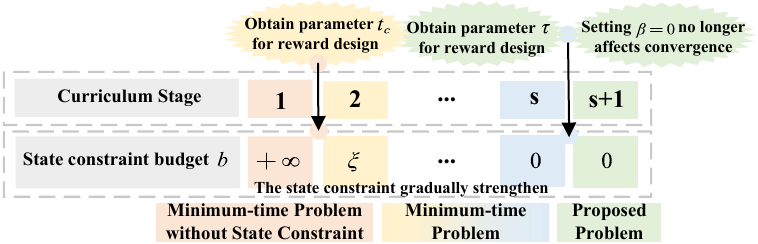}
		\caption{Curriculum learning for reward parameter acquisition}
		\label{fig_cl1}
		\vspace{-5pt} 
	\end{figure}
	
	The multi-stage curriculum is illustrated in Figure \ref{fig_cl1}. Specifically, Stage 1, Stage $s$, and Stage $s+1$ provide the optimal joint policies for the \textit{minimum-time problem} without state constraints (\ref{safe2}), the \textit{minimum-time problem}, and the proposed problem (\ref{eq:P2}), respectively. The joint policies obtained from the first two problems provide essential information for reward design.

	Solving the unconstrained \textit{minimum-time problem} before the constrained version aligns with the CL principle of addressing simpler tasks before more complex ones. Based on this concept, we present a detailed curriculum division, as illustrated in Figure \ref{fig_cl1}, for curriculum learning stages $1$ to $s$. Without loss of generality, we define the feasible state set in (\ref{safe2}) as $C = \{x(t) \in \mathbb{R}^{np} : g[x(t)] \leq 0\}$, with the state constraint budget $b$ set to 0. In the initial stage, which addresses the \textit{minimum-time problem} without state constraints, the constraint budget is considered as $b=+\infty$. In the subsequent stage $j$ ($2 \leq j \leq s$), the state constraints are incrementally enforced with diminishing budget $b=\xi - \frac{(j-2)\xi}{s-2}$. Here, $s$ is the total number of stages, and $\xi$ is the state constraint budget for stage 2, determined based on stage 1's adherence to state constraints. By stage $s$, the budget reaches zero, aligning with (\ref{safe2}). 
	This method initially focuses on terminal constraints and gradually incorporates state constraints, accelerating convergence through two key strategies: 1) streamlining exploration by concentrating on transitions between initial and final states, and 2) allowing agents to learn policies in simpler scenarios with weaker constraints before addressing more complex ones. This approach avoids the inefficiency and potential convergence issues caused by excessive penalties, as discussed in \cite{wang2023cama}.
	
	Moreover, at the convergence of the $s^\text{th}$ curriculum stage, agents obtain a joint policy that reaches the terminal state while satisfying state constraints, eliminating the need for guidance rewards. Thus, setting the guidance reward weight $\beta = 0$ does not affect training convergence, validating the reward settings in Theorem 1.


	\section{Experiments}\label{sec:sim}
	
\begin{figure}[H]
		\centering
		\vspace{-3pt} 
		\includegraphics[width=0.225\textwidth]{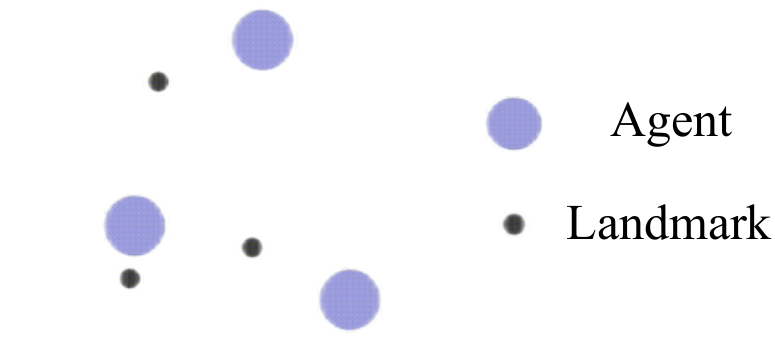}
		\vspace{-8pt}
		\caption{Coverage control task}
		\label{fig:mpe}
	\end{figure}

	We validate our proposed reward design methodology within a multi-agent particle environment \cite{lowe2017multi, hu2024distributed}, a common RL setting, focusing on a coverage control task. 
	As illustrated in Fig. \ref{fig:mpe}, there are three agents (indicated by purple disks) and three landmarks (denoted by black disks) positioned within a two-dimensional space. 
	The terminal constraint is satisfied when the agents complete coverage of the landmarks. This is defined by the condition \(\sum_{i=1}^{3} \min\limits_{k \in\{1,2, 3\}} {d_{ik}} < 0.6\), where \(d_{ik}\) denotes the distance between agent \(i\) and landmark \(k\). The state constraint is designed to prevent collisions among the agents, stipulating that for any two agents \( i \) and \( j \), the distance \( r_{ij} \) between them should satisfy \( r_{ij} > r = 0.3 \), where \( r \) represents the safety distance. We examine two specific forms of the proposed problem (\ref{eq:P2}), which differ in their optimization objectives: 1) \textit{The minimum-time problem;} 2) \textit{The minimum-action problem}: At each time step, an agent has the option to move ``up," ``down," ``left," ``right," or remain ``stationary." We aim to minimize the total number of actions executed by all agents. To achieve this, the cost function is defined as $c(s^t, a^t) = \sum_{i=1}^{3} v_i^t.$ Here, we define
	$v_i^t = 
	\begin{cases} 
		0, & \text{if agent } i \text{ is stationary at time } t\\ 
		1, & \text{if agent } i \text{ moves at time } t
	\end{cases}$. For both problems, we set the guidance function as
	\(  l(s^t, a^t) = 0.5 \sum_{i=1}^{3} \min\limits_{k \in\{1,2,3\}} d_{ik} + 1.35 \).

	We assess the performance using three metrics derived from results in 30 parallel simulation environments. 
	\begin{enumerate}[itemsep=0pt,topsep=0pt,parsep=0pt, leftmargin=*]
		\item  Terminal constraint violation rate \( p_\ell^{m} \) at training step $\ell$: The rate is calculated as \( p_\ell^{m} = \frac{\sum_{k=1}^{30}z^m_{\ell,k}}{30} \), where \(z^m_{\ell,k}\) is a binary indicator of terminal constraint violations in the \( k \)-th parallel environment during the episode corresponding to training step $\ell$. We set \( z^m_{\ell,k} = 1\) if there's a collision or incomplete coverage; otherwise, \( z^m_{\ell,k} = 0 \).
		\item  State constraint violation rate \( p_\ell^{s}\) at training step $\ell$: This rate is denoted as \( p_\ell^{s} = \frac{\sum_{k=1}^{30}z^s_{\ell,k}}{30} \), where \(z^m_{\ell,k}\) is a binary variable indicating whether a state constraint violation occurred in the \( k \)-th parallel environment during the episode corresponding to training step $\ell$. The value of $z^s_{\ell,k}$ is set to 1 if a collision occurs; otherwise, \( z^s_{\ell,k} = 0 \). 
		\item  Optimization objective: For the \textit{minimum-time problem}, this metric at training step $\ell$ is $t^{f}_{\ell}=\frac{t^{f}_{\ell,k}}{30}$, where $t^{f}_{\ell,k}$ is the terminal time in the \( k \)-th parallel environment during the episode corresponding to training step $\ell$. The value of $t^{f}_{\ell,k}$ is set to 50 (twice the maximum time steps per episode) if any constraint is unmet; otherwise, the actual time is used. For the minimum-action problem, the objective for training step $\ell$ is $j_{\ell}=\frac{j_{\ell,k}}{30}$, where $j_{\ell,k}$ denotes the total actions taken in the \( k \)-th parallel environment during the episode corresponding to training step $\ell$. The value of $j_{\ell,k}$ is 75 (the product of the maximum time steps per episode and the maximum movement steps per time step) if any constraint is unmet; otherwise, the actual value is used. 
	\end{enumerate}
	Every 7500 training steps, we sample the terminal constraint violation rate \( p_\ell^{m} \), state constraint violation rate \( p_\ell^{s}\), and optimization objective $t^{f}_{\ell}$ or $j_{\ell}$. Then, we calculate the average of 100 samples around each sampling point, resulting in the average terminal constraint violation rate \( \overline{p_\ell^{m}} \), average state constraint violation rate \( \overline{p_\ell^{s}}\), and average optimization objective $\overline{t^{f}_{\ell}} $ or $\overline{j_{\ell}} $.
	Lower values are preferable for these metrics. 
	
	The experiments are designed to address two key questions: 1). Does our proposed reward design enhance performance compared to existing frameworks? To address this, we conduct a comparative analysis of our reward scheme against established methodologies \cite{lowe2017multi, hu2024distributed}. 2). Does our reward design offer a more efficient solution relative to alternative weighting schemes? To evaluate this, we set the weight \( \alpha \) of the reward \(\mathcal{R}^{a}(s^t, a^t)\) to 10 and adjust the weights of other rewards accordingly. 
	
	To ensure a fair comparison across all scenarios, we employ a consistent set of parameters as follows. Both the critic and policy networks are initialized with a single hidden layer of 64 neurons. Besides, techniques like normalization and ReLU activation are used, as in \cite{yan2022relative}.
	
	\subsection{Experiment for the Minimum-time Problem}
\begin{figure*}
		\centering
		\vspace{-3pt} 
		\includegraphics[width=1\textwidth]{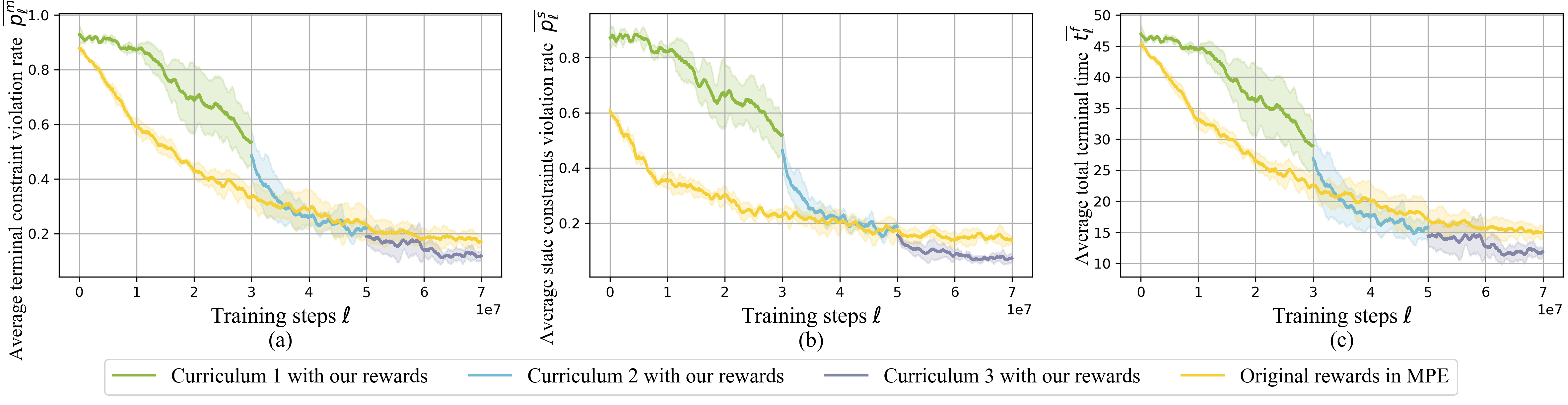}
		\vspace{-8pt}
		\caption{Training under our reward design and the original one \cite{lowe2017multi, hu2024distributed} for the \textit{minimum-time problem} across 3 random seeds}
		\label{fig:reward1}
	\end{figure*}

	We initially focus on the \textit{minimum-time problem}, which corresponds to the proposed problem (\ref{eq:P2}) with a specified optimization metric. To address question 1), we compare our reward design framework with the established one \cite{lowe2017multi, hu2024distributed}. The established framework \cite{lowe2017multi, hu2024distributed} differs from ours in three main aspects: (1) Reward Structure: The reward settings in \cite{lowe2017multi, hu2024distributed} differ from ours; for detailed definitions, refer to \cite{lowe2017multi, hu2024distributed}. (2) Definition of the Terminal step $T_f$: We set $T_f$ to the current time step $t$ if state constraints are violated or terminal conditions are met, whereas \cite{lowe2017multi, hu2024distributed} does not. (3) Curriculum Learning: Our framework incorporates CL, which is not included in \cite{lowe2017multi, hu2024distributed}.
	In terms of our reward framework, the safe distance \( r \), reward parameters \(\alpha\), \(\lambda\), and \(\beta\), and discount factor \(\gamma_m\) for each curriculum stage are summarized in Table \ref{tb:cl1}. To prevent numerical instability, parameters \( \lambda \) and \( \beta \) are configured to approach their bounds according to Theorem 2 and Corollary 1. Fig. \ref{fig:reward1} illustrates the comparative analysis, demonstrating that by the final curriculum stage, our reward design (purple line) surpasses existing methods (yellow line) across all performance metrics.
	
	\begin{table}[htbp]
		\centering
		\fontsize{8pt}{8pt}\selectfont 
		\vspace{-5pt} 
		\caption{Curriculum for the \textit{minimum-time problem}}
		\begin{tabular}{ccccccc}
			\hline
			Stage & Training steps & $r$  &   $\alpha$ &   $\lambda$ & $\beta$  & $\gamma_m$\\
			\hline
			1 & 3$\times10^7$  & 0   &  10  & / & 0.035 & 0.99 \\
			2 & 2$\times10^7$ & 0.25 &  10 & 12.5 & 0.035& 0.99 \\
			3 & 2$\times10^7$  & 0.3 &  10 & 12.5 & 0.035& 0.99 \\
			\hline
		\end{tabular}
		\label{tb:cl1}
		\vspace{-5pt} 
	\end{table}

\begin{figure*}
		\centering
		\vspace{-3pt} 
		\includegraphics[width=1\textwidth]{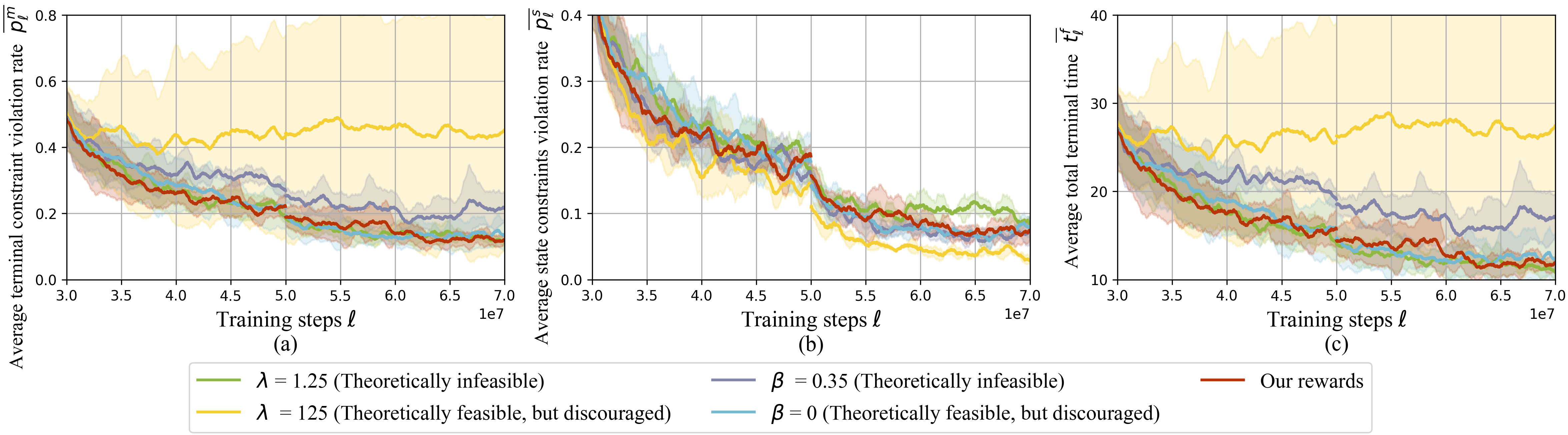}
		\vspace{-8pt}
		\caption{Training under different reward designs for the \textit{minimum-time problem} across 3 random seeds}
		\label{fig:reward2}
	\end{figure*}

	With \( \alpha \) set to 10, we evaluate the performance of reward parameters \( \lambda \) and \( \beta \) that satisfy and are near the bounds of Theorem 2, comparing it to two cases: case 1, where parameter \( \lambda \) or \( \beta \) violating Theorem 2; case 2, where parameter \( \lambda \) or \( \beta \) meeting but not nearing Theorem 2's bound. Given these parameters are set post-curriculum stage 1, our focus is on their effects during stages 2-3.
	The line for case 1 is shown in the first row of the legend in Fig. \ref{fig:reward2}. An insufficient penalty weight \( \lambda \) of 1.25 (0.1 times the value meeting and nearing the bound of Theorem 2 (VMNBT2)) incurs more state constraint breaches, as shown in the green line of Fig. \ref{fig:reward2}(b). An overly large guidance reward weight \( \beta \) of 0.35 (10 times the VMNBT2) results in higher terminal constraint violation rates and longer terminal times due to excessive misguidance, as shown in the purple line of Fig. \ref{fig:reward2}(a) and Fig. \ref{fig:reward2}(c). We then assess case 2, as indicated in the last row of the legend in Fig. \ref{fig:reward2}. A severe penalty weight of \( \lambda = 125 \) (10 times the VMNBT2) leads to increased terminal constraint violations and longer terminal times, despite ensuring state constraint compliance, as shown in the yellow line of Fig. \ref{fig:reward2}(a) and Fig. \ref{fig:reward2}(c). With guidance reward weight \( \beta = 0 \), the average metrics are similar to our method, but with higher variance, as shown in the blue line at the end of training in Fig. \ref{fig:reward2}(a) and Fig. \ref{fig:reward2}(c). This suggests that the absence of guidance rewards may affect robustness. Taken together, our reward scheme demonstrated superior overall performance.

	\subsection{Experiment for the Minimum-action Problem}
	Next, we focus on the \textit{minimum-action problem}, which corresponds to the proposed problem (\ref{eq:P2}) with a specified optimization metric. Solving the \textit{minimum-action problem} requires first solving the \textit{minimum-time problem}, as shown in Fig. \ref{fig_cl1}. Specifically, the first three stages follow the curriculum outlined in Table \ref{tb:cl1} and adhere to the reward and discount factor pair $\langle\mathcal{R}_m, \gamma_m\rangle$ of the \textit{minimum-time problem}, while the fourth stage is defined in Table \ref{tb:cl2}. In our reward design framework, the parameters \( \lambda \), \( \beta \) and \( \mu \) are configured according to Theorem 1, with \( \mu \) and \( \lambda \) positioned near their respective bounds.
	\begin{table}[htbp]
		\centering
		\fontsize{8pt}{8pt}\selectfont 
		\vspace{-5pt} 
		\caption{Curriculum for the \textit{minimum-action problem}}
		\begin{tabular}{cccccccc}
			\hline
			Stage & Training steps & $r$ &   $\alpha$  &   $\lambda$ & $\beta$ & $\mu$ & $\gamma$\\
			\hline
			4 & 4$\times10^7$  & 0.3 &   10 & 10.5 & 0 & -0.12 & 1 \\
			\hline
		\end{tabular}
		\label{tb:cl2}
		\vspace{-5pt} 
	\end{table}

	\begin{figure*}[h]
		\centering
		\vspace{-3pt} 
		\includegraphics[width=1\textwidth]{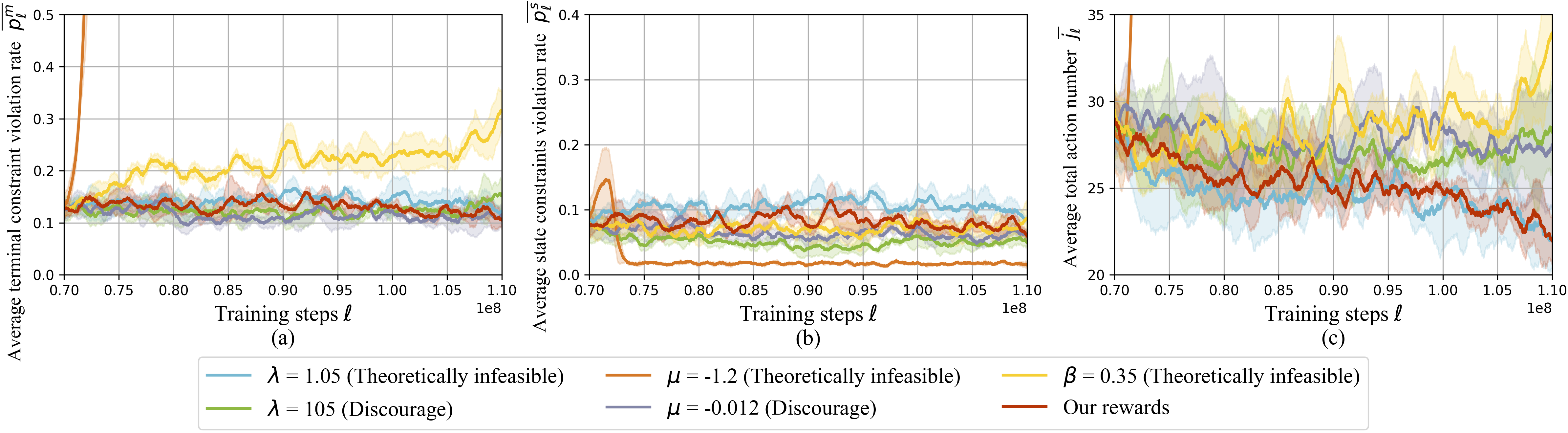}
		\vspace{-8pt}
		\caption{Training under different reward designs for the \textit{minimum-action problem} across 3 random seeds}
		\label{fig:reward3}
	\end{figure*}

	With \( \alpha \) set to 10, we evaluate the performance of \( \lambda \), \( \mu\), and \( \beta \) that meet and closely approach the bounds of Theorem 1, against two cases: case 1, where parameter \( \lambda \), \( \mu\) or \( \beta \) violates Theorem 1; case 2, where parameter \( \lambda \), \( \mu\) or \( \beta \) meets but does not closely approach Theorem 1's bound. Since our reward parameters are established at the end of curriculum stage 3, we concentrate our comparison on stage 4.
	The analysis for case 1 is indicated in the first row of the legend in Fig. \ref{fig:reward3}. An insufficient penalty weight \( \lambda=1.05 \) (0.1 times the value meeting and nearing the bound of Theorem 1 (VMNBT1)) results in increased state constraint violations, as shown in the blue line of Fig. \ref{fig:reward3}(b). The large cost function weight \( \mu =-1.2\) (10 times the VMNBT1) elevates terminal constraint violation rates, as shown in the orange line of Fig. \ref{fig:reward3}(a). This may be due to the movement cost overshadowing the reward for reaching the terminal state, thereby reducing its attractiveness. A large guidance reward weight \( \beta=0.35 \) leads to higher terminal constraint violation rates and total action counts, as shown in the yellow line of Fig. \ref{fig:reward3}(a) and Fig. \ref{fig:reward3}(c).
	We then evaluate case 2, as indicated in the last row of the legend in Fig. \ref{fig:reward3}. Setting the penalty weight \( \lambda = 105\)  (10 times the VMNBT1) increases total action counts despite maintaining state constraint adherence, as shown in the green line of Fig. \ref{fig:reward3}(c). This suggests that severe penalties might cause agents to focus excessively on state constraints to the detriment of other objectives. When the cost function weight \( \mu = -0.012 \) (0.1 times the VMNBT1) is low, the total action count does not decrease, as shown in the purple line of Fig. \ref{fig:reward3}(c). In conclusion, only our reward scheme (red line) and the penalty weight \( \lambda = 1.05 \) (blue line) successfully minimize total actions, as shown in Fig. \ref{fig:reward3}(c). Our approach outperforms the penalty weight \( \lambda = 1.05 \) in state constraint satisfaction, as shown in Fig. \ref{fig:reward3}(b).

	\noindent $\mathbf{Remark\ 1}$. Given that the original multi-agent particle environment does not consider the minimization of total actions, we compare our reward scheme solely with alternative weight configurations.

	\section{Conclusion}\label{sec:con}
	We present a reward design framework for the RL-based multi-agent constrained optimal control problem. Initially, we provide a unified POMDP framework where the final reward is the weighted sum of four components. Subsequently, we establish bounds for the weights of these components based on theoretical proofs, ensuring that the policy maximizing the reward satisfies constraints, minimizes control cost, and avoids numerical instability. Recognizing the need for prior knowledge in the reward setting, we solve two subproblems sequentially, with each informing the reward configuration for the next. We then integrate CL with RL to apply policies from simpler subproblems to more complex ones, accelerating convergence. Experimental results demonstrate the superiority of our approach in constraint satisfaction and control cost minimization.
	Taken together, this paper presents a reward scheme aimed at simplifying the complex reward-tuning process for those working on RL-based optimal control problems. As the current design is tailored for deterministic systems, future research will investigate extending its applicability to systems with stochastic disturbances.
	
	\bibliographystyle{model5-names}     
	\bibliography{reference}       

\begin{thebibliography}{32}
\providecommand{\natexlab}[1]{#1}
\providecommand{\url}[1]{\texttt{#1}}
\expandafter\ifx\csname urlstyle\endcsname\relax
  \providecommand{\doi}[1]{doi: #1}\else
  \providecommand{\doi}{doi: \begingroup \urlstyle{rm}\Url}\fi

\bibitem[Bengio et~al.(2009)Bengio, Louradour, Collobert, and Weston]{cl2009}
Yoshua Bengio, J{\'e}r{\^o}me Louradour, Ronan Collobert, and Jason Weston.
\newblock Curriculum learning.
\newblock In \emph{Proceedings of the 26th annual international conference on
  machine learning}, pages 41--48, 2009.

\bibitem[Bertsekas(2019)]{bertsekas2019reinforcement}
Dimitri Bertsekas.
\newblock \emph{Reinforcement learning and optimal control}, volume~1.
\newblock Athena Scientific, 2019.

\bibitem[Booth et~al.(2023)Booth, Knox, Shah, Niekum, Stone, and
  Allievi]{booth2023perils}
Serena Booth, W~Bradley Knox, Julie Shah, Scott Niekum, Peter Stone, and
  Alessandro Allievi.
\newblock The perils of trial-and-error reward design: misdesign through
  overfitting and invalid task specifications.
\newblock In \emph{Proceedings of the AAAI Conference on Artificial
  Intelligence}, volume~37, pages 5920--5929, 2023.

\bibitem[Dey et~al.(2024)Dey, Dasgupta, and Dey]{P2BPO2024}
Sumanta Dey, Pallab Dasgupta, and Soumyajit Dey.
\newblock P2bpo: Permeable penalty barrier-based policy optimization for safe
  rl.
\newblock In \emph{Proceedings of the AAAI Conference on Artificial
  Intelligence}, volume~38, pages 21029--21036, 2024.

\bibitem[Elbert et~al.(2013)Elbert, Ebbesen, and Guzzella]{Dynamic2013}
Philipp Elbert, Soren Ebbesen, and Lino Guzzella.
\newblock Implementation of dynamic programming for $n$-dimensional optimal
  control problems with final state constraints.
\newblock \emph{IEEE Transactions on Control Systems Technology}, 21\penalty0
  (3):\penalty0 924--931, 2013.

\bibitem[Gregory(2018)]{variations2018}
John Gregory.
\newblock \emph{Constrained optimization in the calculus of variations and
  optimal control theory}.
\newblock Chapman and Hall/CRC, 2018.

\bibitem[Gu et~al.(2024)Gu, Yang, Du, Chen, Walter, Wang, and
  Knoll]{gu2024review}
Shangding Gu, Long Yang, Yali Du, Guang Chen, Florian Walter, Jun Wang, and
  Alois Knoll.
\newblock A review of safe reinforcement learning: Methods, theories and
  applications.
\newblock \emph{IEEE Transactions on Pattern Analysis and Machine
  Intelligence}, 2024.

\bibitem[Han et~al.(2021)Han, Tian, Zhang, Wang, and Pan]{han2021reinforcement}
Minghao Han, Yuan Tian, Lixian Zhang, Jun Wang, and Wei Pan.
\newblock Reinforcement learning control of constrained dynamic systems with
  uniformly ultimate boundedness stability guarantee.
\newblock \emph{Automatica}, 129:\penalty0 109689, 2021.

\bibitem[Hayes et~al.(2022)Hayes, R{\u{a}}dulescu, Bargiacchi,
  K{\"a}llstr{\"o}m, Macfarlane, Reymond, Verstraeten, Zintgraf, Dazeley,
  Heintz, et~al.]{hayes2022practical}
Conor~F Hayes, Roxana R{\u{a}}dulescu, Eugenio Bargiacchi, Johan
  K{\"a}llstr{\"o}m, Matthew Macfarlane, Mathieu Reymond, Timothy Verstraeten,
  Luisa~M Zintgraf, Richard Dazeley, Fredrik Heintz, et~al.
\newblock A practical guide to multi-objective reinforcement learning and
  planning.
\newblock \emph{Autonomous Agents and Multi-Agent Systems}, 36\penalty0
  (1):\penalty0 26, 2022.

\bibitem[He et~al.(2023)He, Dong, Song, and Sun]{MPtime}
Zichen He, Lu~Dong, Chunwei Song, and Changyin Sun.
\newblock Multiagent soft actor-critic based hybrid motion planner for mobile
  robots.
\newblock \emph{IEEE Transactions on Neural Networks and Learning Systems},
  34\penalty0 (12):\penalty0 10980--10992, 2023.

\bibitem[Hu et~al.(2024)Hu, Fu, Wen, Lv, and Ren]{hu2024distributed}
Yifan Hu, Junjie Fu, Guanghui Wen, Yuezu Lv, and Wei Ren.
\newblock Distributed entropy-regularized multi-agent reinforcement learning
  with policy consensus.
\newblock \emph{Automatica}, 164:\penalty0 111652, 2024.

\bibitem[Huang et~al.(2022)Huang, Wang, Zhou, Zhang, and Lin]{huang2022reward}
Changxin Huang, Guangrun Wang, Zhibo Zhou, Ronghui Zhang, and Liang Lin.
\newblock Reward-adaptive reinforcement learning: Dynamic policy gradient
  optimization for bipedal locomotion.
\newblock \emph{IEEE transactions on pattern analysis and machine
  intelligence}, 45\penalty0 (6):\penalty0 7686--7695, 2022.

\bibitem[Kim et~al.(2024)Kim, Oh, Lee, Choi, Ji, Jung, Youm, and
  Hwangbo]{IPO2024}
Yunho Kim, Hyunsik Oh, Jeonghyun Lee, Jinhyeok Choi, Gwanghyeon Ji, Moonkyu
  Jung, Donghoon Youm, and Jemin Hwangbo.
\newblock Not only rewards but also constraints: Applications on legged robot
  locomotion.
\newblock \emph{IEEE Transactions on Robotics}, 40:\penalty0 2984--3003, 2024.

\bibitem[Lewis et~al.(2012)Lewis, Vrabie, and Syrmos]{lewis2012optimal}
Frank~L Lewis, Draguna Vrabie, and Vassilis~L Syrmos.
\newblock \emph{Optimal control}.
\newblock John Wiley \& Sons, 2012.

\bibitem[Li et~al.(2022)Li, Li, Zeng, Rong, and Zhang]{MPcontrolcost}
Bin Li, Qingliang Li, Yong Zeng, Yue Rong, and Rui Zhang.
\newblock 3d trajectory optimization for energy-efficient uav communication: A
  control design perspective.
\newblock \emph{IEEE Transactions on Wireless Communications}, 21\penalty0
  (6):\penalty0 4579--4593, 2022.

\bibitem[Liu et~al.(2022)Liu, Li, and Ji]{FR}
Gaoqi Liu, Bin Li, and Yuandong Ji.
\newblock A modified hp-adaptive pseudospectral method for multi-uav formation
  reconfiguration.
\newblock \emph{ISA Transactions}, 129:\penalty0 217--229, 2022.

\bibitem[Liu et~al.(2020)Liu, Ding, and Liu]{IPO2020}
Yongshuai Liu, Jiaxin Ding, and Xin Liu.
\newblock Ipo: Interior-point policy optimization under constraints.
\newblock In \emph{Proceedings of the AAAI conference on artificial
  intelligence}, volume~34, pages 4940--4947, 2020.

\bibitem[Lowe et~al.(2017)Lowe, Wu, Tamar, Harb, Abbeel, and
  Mordatch]{lowe2017multi}
Ryan Lowe, Yi~Wu, Aviv Tamar, Jean Harb, Pieter Abbeel, and Igor Mordatch.
\newblock Multi-agent actor-critic for mixed cooperative-competitive
  environments.
\newblock \emph{Neural Information Processing Systems}, 2017.

\bibitem[Moradian and Kia(2021)]{cluster2021}
Hossein Moradian and Solmaz~S Kia.
\newblock Cluster-based distributed augmented lagrangian algorithm for a class
  of constrained convex optimization problems.
\newblock \emph{Automatica}, 129:\penalty0 109608, 2021.

\bibitem[Ng et~al.(1999)Ng, Harada, and Russell]{ng1999policy}
Andrew~Y Ng, Daishi Harada, and Stuart Russell.
\newblock Policy invariance under reward transformations: Theory and
  application to reward shaping.
\newblock In \emph{Icml}, volume~99, pages 278--287, 1999.

\bibitem[Paruchuri and Chatterjee(2019)]{Pontryagin2019}
Pradyumna Paruchuri and Debasish Chatterjee.
\newblock Discrete time pontryagin maximum principle under
  state-action-frequency constraints.
\newblock \emph{IEEE Transactions on Automatic Control}, 64\penalty0
  (10):\penalty0 4202--4208, 2019.

\bibitem[Ravaioli et~al.(2022)Ravaioli, Cunningham, McCarroll, Gangal, Dunlap,
  and Hobbs]{SR}
Umberto~J. Ravaioli, James Cunningham, John McCarroll, Vardaan Gangal, Kyle
  Dunlap, and Kerianne~L. Hobbs.
\newblock Safe reinforcement learning benchmark environments for aerospace
  control systems.
\newblock In \emph{2022 IEEE Aerospace Conference (AERO)}, pages 1--20, 2022.

\bibitem[Rizvi and Boyle(2025)]{10812765}
Danish Rizvi and David Boyle.
\newblock Multi-agent reinforcement learning with action masking for
  uav-enabled mobile communications.
\newblock \emph{IEEE Transactions on Machine Learning in Communications and
  Networking}, 3:\penalty0 117--132, 2025.
\newblock \doi{10.1109/TMLCN.2024.3521876}.

\bibitem[Safaoui et~al.(2024)Safaoui, Vinod, Chakrabarty, Quirynen, Yoshikawa,
  and Cairano]{safaoui2024safe}
Sleiman Safaoui, Abraham~P. Vinod, Ankush Chakrabarty, Rien Quirynen, Nobuyuki
  Yoshikawa, and Stefano~Di Cairano.
\newblock Safe multiagent motion planning under uncertainty for drones using
  filtered reinforcement learning.
\newblock \emph{IEEE Transactions on Robotics}, 40:\penalty0 2529--2542, 2024.

\bibitem[Sootla et~al.(2022)Sootla, Cowen-Rivers, Jafferjee, Wang, Mguni, Wang,
  and Ammar]{2022saute}
Aivar Sootla, Alexander~I Cowen-Rivers, Taher Jafferjee, Ziyan Wang, David~H
  Mguni, Jun Wang, and Haitham Ammar.
\newblock Saut{\'e} rl: Almost surely safe reinforcement learning using state
  augmentation.
\newblock In \emph{International Conference on Machine Learning}, pages
  20423--20443. PMLR, 2022.

\bibitem[Sutton and Barto(2018)]{Reinforcement}
R.~S. Sutton and A.~G. Barto.
\newblock \emph{Reinforcement learning: An introduction}.
\newblock MIT press, 2018.

\bibitem[Tessler et~al.(2019)Tessler, Mankowitz, and Mannor]{tessler2018reward}
Chen Tessler, Daniel~J. Mankowitz, and Shie Mannor.
\newblock Reward constrained policy optimization.
\newblock In \emph{International Conference on Learning Representations}, 2019.

\bibitem[Tipaldi et~al.(2022)Tipaldi, Iervolino, and Massenio]{PL}
Massimo Tipaldi, Raffaele Iervolino, and Paolo~Roberto Massenio.
\newblock Reinforcement learning in spacecraft control applications: Advances,
  prospects, and challenges.
\newblock \emph{Annual Reviews in Control}, 54:\penalty0 1--23, 2022.

\bibitem[Wang et~al.(2023)Wang, Du, Sootla, Ammar, and Wang]{wang2023cama}
Ziyan Wang, Yali Du, Aivar Sootla, Haitham~Bou Ammar, and Jun Wang.
\newblock {CAMA}: A new framework for safe multi-agent reinforcement learning
  using constraint augmentation, 2023.
\newblock URL \url{https://openreview.net/forum?id=jK02XX9ZpJkt}.

\bibitem[Yan et~al.(2022)Yan, Li, Qiu, Qiu, Wang, Wang, and
  Shen]{yan2022relative}
Yuzi Yan, Xiaoxiang Li, Xinyou Qiu, Jiantao Qiu, Jian Wang, Yu~Wang, and Yuan
  Shen.
\newblock Relative distributed formation and obstacle avoidance with
  multi-agent reinforcement learning.
\newblock In \emph{2022 international conference on robotics and automation
  (ICRA)}, pages 1661--1667. IEEE, 2022.

\bibitem[Yu et~al.(2022)Yu, Velu, Vinitsky, Gao, Wang, Bayen, and
  Wu]{mappo2022}
Chao Yu, Akash Velu, Eugene Vinitsky, Jiaxuan Gao, Yu~Wang, Alexandre Bayen,
  and Yi~Wu.
\newblock The surprising effectiveness of ppo in cooperative multi-agent games.
\newblock \emph{Advances in Neural Information Processing Systems},
  35:\penalty0 24611--24624, 2022.

\bibitem[Zhang et~al.(2022)Zhang, Shen, Yang, Chen, Wang, Yuan, and
  Tao]{P3O2022}
Linrui Zhang, Li~Shen, Long Yang, Shixiang Chen, Xueqian Wang, Bo~Yuan, and
  Dacheng Tao.
\newblock Penalized proximal policy optimization for safe reinforcement
  learning.
\newblock In \emph{International Joint Conference on Artificial Intelligence},
  2022.

\end{thebibliography}
	
	\appendix
	
	\setstretch{0.975}
	\setlength{\jot}{0pt} 
	
	\section{Proof of Theorem 1}\label{sec:T2}
	Due to the comprehensive nature of the proof, we have organized it into 5 steps to enhance readability, with the crucial content of each step emphasized in $\mathbf{bold}$.
	
	$\mathbf{\noindent Step\ 1.\ Decompose}$ $J(s^0, \pi)$$\mathbf{.}$ Taking into account the noise-free system model (\ref{motion2}), we acquire
	\begin{equation}
		J(s^0, \pi) = \mathbb{E}\left[\sum_{t=0}^{T_f}\gamma^{t}\mathcal{R}(s^t, a^t)\right] 
		=\sum_{t=0}^{T_f}\gamma^{t}\mathcal{R}(s^t, a^t).
		\label{eq:J2*}
	\end{equation}
	Substituting (\ref{eq:gamma}) into (\ref{eq:J2*}), we have
	\begin{align}
		J(s^0, \pi) =\sum_{t=0}^{T_f}\mathcal{R}(s^t, a^t). \label{eq:J2gamma1*}
	\end{align}
	Substituting (\ref{eq:reward2}) into (\ref{eq:J2gamma1*}), we obtain:
	\begin{equation}
		\begin{aligned}
			J(s^0, \pi) =&\sum_{t=0}^{T_f}\Big( \alpha\mathcal{R}^{a}(s^t, a^t) + \beta\mathcal{R}^{g}(s^t, a^t)  \\
			&+ \lambda\mathcal{R}^{p}(s^t, a^t) + \mu\mathcal{R}^{c}(s^t, a^t)\Big).
		\end{aligned}
		\label{eq:2J2}
	\end{equation}
	Defining
	\begin{subequations}
		\begin{align}
			&\label{eq:Ja}J^{a}(s^0,\pi)=
			\sum_{t=0}^{T_f}\alpha\mathcal{R}^{a}(s^t, a^t).\\
			&\label{eq:Jn}J^{g}(s^0,\pi)=
			\sum_{t=0}^{T_f}\beta\mathcal{R}^{g}(s^t, a^t).\\
			&\label{eq:Jc}J^{p}(s^0,\pi)=
			\sum_{t=0}^{T_f}\lambda\mathcal{R}^{p}(s^t, a^t).\\
			&\label{eq:Jcost2}J^{c}(s^0,\pi)=
			\sum_{t=0}^{T_f}\mu\mathcal{R}^{c}(s^t, a^t).
		\end{align}
	\end{subequations}
	Substituting (\ref{eq:Ja})-(\ref{eq:Jcost2}) into (\ref{eq:2J2}), we get
	\begin{equation}
		J(s^0, \pi) =J^{a}(s^0,\pi)+J^{g}(s^0,\pi)+J^{p}(s^0,\pi)+J^{c}(s^0,\pi). \label{eq:Jall*}
	\end{equation}
	When we substitute $\beta = 0$ into (\ref{eq:Jn}), we derive
	\begin{equation}
		J^{g}(s^0,\theta)=0.
		\label{eqJn0}
	\end{equation}
	By substituting (\ref{eqJn0}) into (\ref{eq:Jall*}), we obtain
	\begin{align}
		J(s^0, \pi) &=J^{a}(s^0,\theta)+J^p(s^0,\theta)+J^{c}(s^0,\theta). \label{eq:Jall1*}
	\end{align}
	
	$\mathbf{\noindent Step\ 2.\ Classify\ policies.}$ We categorize joint policies into four main sets $\Pi^i,i = 0,1,2,3$ as shown in Table \ref{table1}.
	
	\begin{table}[htbp]
		\centering  
		\vspace{-5pt} 
		\caption{Policy classification}  
		\fontsize{7.5pt}{6.5pt}\selectfont 
		\label{table1}  
		\begin{tabular}{|c|c|c|}  
			\hline 
			&Satisfy terminal constraint&Satisfy state constraint\\  
			\hline
			$\Pi^0$&$\times$&$\times$ \\
			\hline
			$\Pi^1$&$\times$&$\checkmark$ \\
			\hline
			$\Pi^2$&$\checkmark$&$\times$ \\
			\hline
			$\Pi^3$&$\checkmark$&$\checkmark$ \\
			\hline
		\end{tabular}
		\vspace{-5pt} 
	\end{table}

	$\mathbf{\noindent Step\ 3.\ Analyze\ the\ components'\ magnitudes\ of}$ $J(s^0, \pi)$ $\mathbf{in\ (\ref{eq:Jall1*})}.$
	To analyze the magnitude of $J(s^0,\pi^{i})$, where $\pi^{i}\in \Pi^i, i = 0,1,2,3$, we first examine the individual components of $J(s^0,\pi^{i})$, namely, $J^{a}(s^0,\pi^{i})$, $J^p(s^0,\pi^{i})$ and $J^{c}(s^0,\pi^{i})$.
	
	Firstly, based on the definition of the cost function \( c:\mathbb{R}^{np} \times \mathbb{R}^{nq}\rightarrow \mathbb{R}^{+}\) and (\ref{eq:rewardc}), the inequality
	\( \mathcal{R}^{c}(s^t, a^t) \geq 0 \) holds. Substituting this inequality and \( \mu \leq 0 \) into equation (\ref{eq:Jcost2}), we obtain:
	\begin{equation}
		J^{c}(s^0,{\pi_{i}})\leq 0.
		\label{eqeb2}
	\end{equation}
	
	Next, let's consider the magnitude of $J^{a}(s^0,\pi_{i})$. Substitute (\ref{eq:arreq}) into (\ref{eq:Ja}) and consider the definition of the terminal time step $T_f$, we derive that
	\begin{equation}
		\begin{aligned}
			J^{a}(s^0,\pi_{i})=
			\left\{
			\begin{array}{ll}
				\alpha, &
				\text{if the terminal constraint is satisfied}, \\ 
				0, &\text{else.}
			\end{array}
			\right.
		\end{aligned}
		\label{eq:Jaa}
	\end{equation}
	
	Finally, let's examine the magnitude of $J^p(s^0,\pi_{i})$. Substitute (\ref{eq:safeeq}) into (\ref{eq:Jc}), we obtain	
	\begin{equation}
		\begin{aligned}		
			J^p(s^0,\pi_{i})=		\left\{		\begin{array}{ll}			-\lambda, &			\text{if any state constraint is violated}, \\ 			0, &\text{else.}		\end{array}		\right.	
		\end{aligned}	
		\label{eq:JL3}
	\end{equation}
	
	In conclusion, we have determined the magnitudes of the individual components of $J(s^0,\pi_{i})$ as shown in (\ref{eqJn0}), (\ref{eqeb2}), (\ref{eq:Jaa}), and (\ref{eq:JL3}). Next, based on the relative magnitudes of these components, we will analyze the returns associated with each joint policy $J(s^0,\pi_{i}), i = 0,1,2,3$.
	
	$\mathbf{Step\ 4.\ Compare\ the\ magnitude\ of\ returns\ under}$ $\mathbf{ each\ policy}$ $J(s^0,\pi_{i}), i = 0,1,2,3.$
	Initially, for all $\pi^{0}\in\Pi^{0}$, we evaluate the magnitude of $J(s^0,\pi^{0})$. As per (\ref{eq:Jall1*}), we have
	\begin{align}
		J(s^0,\pi^{0}) &={J}_i^{c}(s^0,\pi^{0})+J^{a}(s^0,\pi^{0})+J^p(s^0,\pi^{0}). \label{eq:Jall10*}
	\end{align}
	According to the definition of policy set $\Pi^{0}$, it is known that the agent following it has not yet satisfied the terminal constraint and state constraint. Substituting these conditions into (\ref{eqeb2}), (\ref{eq:Jaa}), and (\ref{eq:JL3}), and considering Theorem 1's conditions $\lambda>\alpha$ and $\alpha>0$, we obtain
	\begin{equation}
		\left\{\begin{array}{ll}
			&J^{c}(s^0,\pi^{0})\leq 0,\\
			&J^{a}(s^0,\pi^{0})= 0,\\
			&J^p(s^0,\pi^{0})=-\lambda<-\alpha<0.
		\end{array}\right.
		\label{eqJpi0}
	\end{equation}
	Through substitution of equation (\ref{eqJpi0}) into equation (\ref{eq:Jall10*}), we conclude that
	\begin{align}
		J(s^0,\pi^{0}) < 0. \label{eqJpi0*}
	\end{align}
	
	Subsequently, for all $\pi^{1}\in\Pi^{1}$, we analyze the magnitude of $J(s^0,\pi^{1})$. According to equation (\ref{eq:Jall1*}), we obtain
	\begin{align}
		J(s^0,\pi^{1}) &={J}_i^{c}(s^0,\pi^{1})+J^{a}(s^0,\pi^{1})+J^p(s^0,\pi^{1}). \label{eq:Jall11*}
	\end{align}
	According to the definition of policy $\Pi^{1}$, it is known that the agent following it has satisfied the state constraint but violated the terminal constraint. Substituting these conditions into (\ref{eqeb2}), (\ref{eq:Jaa}), and (\ref{eq:JL3}), we obtain
	\begin{equation}
		\left\{\begin{array}{ll}
			&J^{c}(s^0,\pi^{1})\leq 0,\\
			&J^{a}(s^0,\pi^{1})= 0,\\
			&J^p(s^0,\pi^{1})=0.
		\end{array}\right.
		\label{eqJpi1}
	\end{equation}
	Replace (\ref{eqJpi1}) into (\ref{eq:Jall11*}), resulting in
	\begin{align}
		J(s^0,\pi^{1}) \leq 0. \label{eqJpi1*}
	\end{align}
	
	In the following, for all $\pi^{2}\in\Pi^{2}$, we analyze the magnitude of $J(s^0,\pi^{2})$. As indicated by (\ref{eq:Jall1*}), we have
	\begin{align}
		J(s^0,\pi^{2}) &=J^{c}(s^0,\pi^{2})+J^{a}(s^0,\pi^{2})+J^p(s^0,\pi^{2}). \label{eq:Jall12*}
	\end{align}
	By the definition of policy set $\Pi^{2}$, it is established that the agent following it has satisfied terminal constraint but violated state constraints. By incorporating these conditions into (\ref{eqeb2}), (\ref{eq:Jaa}), and (\ref{eq:JL3}), we find that
	\begin{equation}
		\left\{\begin{array}{ll}
			&J^{c}(s^0,\pi^{2})\leq 0,\\
			&J^{a}(s^0,\pi^{2})= \alpha,\\
			&J^p(s^0,\pi^{2})=-\lambda.
		\end{array}\right.
		\label{eqJpi211}
	\end{equation}
	By incorporating (\ref{eqJpi211}) into (\ref{eq:Jall12*}), we have
	\begin{align}
		J(s^0,\pi^{2}) \leq \alpha-\lambda. \label{eqJpi266}
	\end{align}
	Then, by substituting $\lambda > \alpha$ in (\ref{eqpenbeta1Pphi}) into (\ref{eqJpi266}), we obtain
	\begin{equation}
		J(s^0,\pi^{2}) <0. \label{eqJpi277}
	\end{equation}
	
	Lastly, for all $\pi^{3}\in\Pi^{3}$, we analyze the magnitude of $J(s^0,\pi^{3})$. According to (\ref{eq:Jall1*}), we have
	\begin{align}
		J(s^0,\pi^{3}) &=J^{c}(s^0,\pi^{3})+J^{a}(s^0,\pi^{3})+J_m^p(s^0,\pi^{3}). \label{eq:Jall13*}
	\end{align}
	According to the definition of policy set $\Pi^{3}$, it is known that the agent following it has satisfied the terminal constraint and the state constraint. Incorporating these conditions into (\ref{eq:Jaa}) and (\ref{eq:JL3}), we have
	\begin{equation}
		\left\{\begin{array}{ll}
			&J^{a}(s^0,\pi^{3})= \alpha,\\
			&J_m^p(s^0,\pi^{3})=0.
		\end{array}\right.
		\label{eqJpi31}
	\end{equation}
	Substitute (\ref{eqJpi31}) into (\ref{eq:Jall13*}), we arrive at
	\begin{align}
		J(s^0,\pi^{3}) = J^{c}(s^0,\pi^{3})+\alpha. \label{eqJpi32}
	\end{align}
	
	The estimation of the magnitude of $J^{c}(s^0,\pi^{3})$ is displaced below. Based on the definition of policy $\tilde\pi$, it can be known that $\tilde\pi\in\Pi^{3}$. Substitute (\ref{eqtua}) into (\ref{eq:Jcost2}) and take into account that $\mu<0$,  we can conclude that
	\begin{equation}
		\begin{aligned}
			J^{c}(s^0,\tilde\pi)\geq \mu \tau,\ \tilde\pi\in \Pi^3.
		\end{aligned}
		\label{eq:Jeb322}
	\end{equation}
	Substitute (\ref{eq:Jeb322}) into (\ref{eqJpi32}), we have
	\begin{align}
		J(s^0,\tilde\pi) \geq \mu \tau+\alpha,\ \tilde\pi\in \Pi^{3}. \label{eqJpi333}
	\end{align}
	Substitute $\mu> - \frac{\alpha}{\tau}$ in (\ref{eqpenbeta1Pphi}) into (\ref{eqJpi333}), we find that
	\begin{align}
		J(s^0,\tilde\pi) > -\alpha+\alpha=0,\ \tilde\pi\in  \Pi^{3}. \label{eqJpi344}
	\end{align}
	
	To sum up, according to (\ref{eqJpi0*}), (\ref{eqJpi1*}) and (\ref{eqJpi277}), for all $\pi^{0,1,2}\in \Pi^0\cup\Pi^1\cup\Pi^2$, it holds that  $J(s^0,\pi^{0,1,2})\leq 0$. According to (\ref{eqJpi344}), there exist $\tilde\pi \in \Pi^{3} $ such that $	J(s^0,\tilde\pi) > 0$. Thus, we derive
	\begin{align}
		J(s^0,\tilde\pi) >J(s^0,\pi^{0,1,2}),\forall \pi^{0,1,2}\in \Pi^0\cup\Pi^1\cup\Pi^2. \label{eqJpi3666}
	\end{align}
	
	$\mathbf{Step\ 5.\ Prove\ that\ the\ policy}$ $\pi^{3*}\in \Pi^3$ $\mathbf{that\ satisfies}$ $J^{c}(s^0,\pi^{3*})\geq J^{c}(s^0,\pi^{3}),\forall \pi^{3}\in\Pi^{3}$ $\mathbf{is\ the\ one\ with}$ $\mathbf{the\  maximum\ return.}$
	
	According to inequality $J^{c}(s^0,\pi^{3*})\geq J^{c}(s^0,\pi^{3}),\forall \pi^{3}\in\Pi^{3}$, we can derive that
	\begin{align}
		J^{c}(s^0,\pi^{3*})\geq J^{c}(s^0,\tilde\pi). \label{eqJpi3best1}
	\end{align}
	Substitute equation (\ref{eqJpi3best1}) into equation (\ref{eqJpi32}), we get
	\begin{align}
		J(s^0,\pi^{3*})\geq J(s^0,\tilde\pi). \label{eqJpi3best2}
	\end{align}
	Insert (\ref{eqJpi3best2}) into (\ref{eqJpi3666}), we have
	\begin{align}
		J(s^0,\pi^{3*}) >J(s^0,\pi^{0,1,2}),\forall \pi^{0,1,2}\in \Pi^0\cup\Pi^1\cup\Pi^2. \label{eqJpi3best3}
	\end{align}
	
	Besides, in keeping with equation $J^{c}(s^0,\pi^{3*})\geq J^{c}(s^0,\pi^{3}),\forall \pi^{3}\in\Pi^{3}$ and (\ref{eqJpi32}), we have
	\begin{align}
		J(s^0,\pi^{3*})\geq J(s^0,\pi^{3}),\forall \pi^{3}\in\Pi^{3} \label{eqJpi3best4}
	\end{align}
	Following (\ref{eqJpi3best3}) and (\ref{eqJpi3best4}), we can conclude that for all $ \pi^{0,1,2,3}\in \Pi^0\cup\Pi^1\cup\Pi^2\cup\Pi^3$, if $J^{c}(s^0,\pi^{3*})> J^{c}(s^0,\pi^{3}),\forall \pi^3\in\Pi^{3}$ then $J(s^0,\pi^{3*})\geq J(s^0,\pi^{0,1,2,3})$. 
	In conclusion, it can be summarized that the policy $\pi^{3*}$, which is the optimal solution to the problem (\ref{eq:P2}), is\ the\ one\ with\ the maximum\ return.

	\section{Proof of Theorem 2}\label{sec:T1}
	Due to the extensive nature of the proof, for improved readability, we have divided it into 9 steps, with the key content of each step highlighted in $\mathbf{bold}$.
	
	$\mathbf{Step\ 1.\ Decompose}$ $J(s^0, \pi)$$\mathbf{.}$ Considering the noise-free system model (\ref{motion2}), we obtain
	\begin{equation}
		J(s^0, \pi) = \mathbb{E}\left[\sum_{t=0}^{T_f}\gamma_m^{t}\mathcal{R}_m(s^t, a^t)\right]=\sum_{t=0}^{T_f}\gamma_m^{t}\mathcal{R}_m(s^t, a^t).
		\label{eq:J21}
	\end{equation}
	Substituting (\ref{eq:rewardt}) into (\ref{eq:J21}), we obtain:
	\begin{equation}
		J(s^0, \pi) =\sum_{t=0}^{T_f}\gamma_m^{t}\Big( \alpha\mathcal{R}^{a}(s^t, a^t) + \beta\mathcal{R}^{g}(s^t, a^t) +  \lambda\mathcal{R}^{p}(s^t, a^t)\Big).
		\label{eq:J2}
	\end{equation}
	Defining
	\begin{subequations}
		\begin{align}
			&\label{eq:Ja2}J^{a}_m(s^0,\pi)=
			\sum_{t=0}^{T_f}\alpha\gamma_m^{t}\mathcal{R}^{a}(s^t, a^t).\\
			&\label{eq:Jn2}J^{g}_m(s^0,\pi)=
			\sum_{t=0}^{T_f}\beta\gamma_m^{t}\mathcal{R}^{g}(s^t, a^t).\\
			&\label{eq:Jc2}J^{p}_m(s^0,\pi)=
			\sum_{t=0}^{T_f}\lambda\gamma_m^{t}\mathcal{R}^{p}(s^t, a^t).
		\end{align}
	\end{subequations}
	Substituting (\ref{eq:Ja2})-(\ref{eq:Jc2}) into  (\ref{eq:J2}), we get
	\begin{align}
		J(s^0, \pi) &=J^{a}_m(s^0,\pi)+J^{g}_m(s^0,\pi)+J^{p}_m(s^0,\pi). \label{eq:Jall}
	\end{align}
	
	$\mathbf{Step\ 2.\ Calculate\ the\ upper\ bound\ of}$ $|J^{g}_m(s^0,\pi)|$$\mathbf{.}$ Substituting (\ref{eq:naveq}) into (\ref{eq:Jn2}), we have
	\begin{align}\label{eqJg}
		\big|J^{g}_m(s^0,\pi)\big| &=\big| \sum_{t=0}^{T_f}\gamma_m^{t}\beta l(s^t, a^t) \big|\\\notag
		&\leq \big| \sum_{t=0}^{T_f}\beta\gamma_m^{t} \big|\big|l(s^t, a^t) \big|\\\notag
		&= \big|  \beta \frac{1 - \gamma_m^{T_f + 1}}{1 - \gamma_m} \big|\big|l(s^t, a^t) \big|\notag.
	\end{align}
	Substituting the inequality $|l(s^t, a^t)| < \rho$ from Assumption 1 into (\ref{eqJg}), we obtain
	\begin{equation}\label{eqJg1}
		\big|J^{g}_m(s^0,\pi)\big| < \rho   \beta \frac{1 - \gamma_m^{T_f + 1}}{1 - \gamma_m}.
	\end{equation}
	According to the definition of the terminal step \( T_f \), once the current time step satisfies \( t = t_{\max} - 1 \), we define \( T_f = t \). Thus, it follows that the inequality \( T_f \leq t_{\max} - 1 \) holds. Substituting this inequality into (\ref{eqJg1}), we get
	\begin{equation}
		\big|J^{g}_m(s^0,\pi)\big|< \rho   \beta \frac{1 - \gamma_m^{t_{\max}}}{1 - \gamma_m}.
	\end{equation}
	Taking into account $\beta<\frac{\alpha\gamma_m^{t_{\max}}(1 - \gamma_m)^2}{2\rho(1 - \gamma_m^{t_{\max}})}$, we can obtain the following inequality
	\begin{equation}
		\big|J^g_m(s^0,\theta)\big|  <\frac{\alpha\gamma_m^{t_{\max}}(1 - \gamma_m)} {2}.
		\label{eqjnupper}
	\end{equation}
	
	$\mathbf{Step\ 3.\ Classify\ policies\  and\   explore\  the\   returns}$\\  
	$\mathbf{relationships.}$ Firstly, we categorize the joint policies into 4 major sets $\Pi^i,i = 0,1,2,3$, as shown in Table \ref{table1}.
	In the following, we reveal that for all $\pi^3\in\Pi^3$, and for all $\pi^{0,1,2}\in \Pi^0\cup\Pi^1\cup \Pi^2$, it holds that $J(s^0,\pi^{3})>J(s^0,\pi^{0,1,2})$. 
	
	$\mathbf{Step\ 4.\ Prove\ that}$ $J(s^0,\pi^{3})>J(s^0,\pi^0)$, $\mathbf{for\ all}$ $\pi^{3}\in\Pi^{3}$ $\mathbf{and}$ $\pi^{0}\in\Pi^{0}.$
	First, let's consider $J(s^0,\pi^{0})$. According to the definition of policy set $\Pi^{0}$ and rewards (\ref{eq:arreq}) and (\ref{eq:safeeq}), for all $\Pi_{0}\in \boldsymbol{\Pi0}$, we have
	\begin{equation}
		\left\{\begin{array}{ll}
			&J_m^{p}(s^0,\pi^0)<0,\\
			&J_m^{a}(s^0,\pi^0)=0.
		\end{array}\right.
		\label{eqJca}
	\end{equation}
	Replace (\ref{eqJca}) into (\ref{eq:Jall}), we have
	\begin{align}
		J(s^0,\pi^0) &<J_m^{g}(s^0,\pi^0). \label{eq:pi0}
	\end{align}
	Next, let's analyze the magnitude of $J(s^0,\pi^{3})$. According to the definition of policy set $\Pi^3$ and rewards (\ref{eq:arreq}) and (\ref{eq:safeeq}), for all $\pi^{3}\in\Pi^{3}$, it leads to
	\begin{equation}
		\left\{\begin{array}{ll}
			&J_m^{p}(s^0,\pi^{3})=0,\\
			&J_m^{a}(s^0,\pi^{3})>\alpha\gamma_m^{t_{\max}}.
		\end{array}\right.
		\label{eqJca3}
	\end{equation}
	Through substitution of (\ref{eqJca3}) into (\ref{eq:Jall}), we obtain
	\begin{align}
		J(s^0,\pi^{3}) & >\alpha\gamma_m^{t_{\max}}+J_m^{g}(s^0,\pi^{3}). \label{eq:pi3}
	\end{align}
	According to (\ref{eq:pi0}) and (\ref{eq:pi3}), we have 
	\begin{equation}
		\label{eq:pi3-1}
		\begin{aligned}
			&J(s^0,\pi^{3})-J(s^0,\pi^0) \\ >&\alpha\gamma_m^{t_{\max}}+J_m^{g}(s^0,\pi^{3})-J_m^{g}(s^0,\pi^0).
		\end{aligned}
	\end{equation}
	In accordance with (\ref{eqjnupper}), one has
	\begin{equation}
		J_m^{g}(s^0,\pi^{3})-J_m^{g}(s^0,\pi^0)>-\alpha\gamma_m^{t_{\max}}(1 - \gamma_m).
		\label{eq:JNmaxp}
	\end{equation}
	Incorporating (\ref{eq:JNmaxp}) into (\ref{eq:pi3-1}), we have
	\begin{equation}
		\begin{aligned}
			J(s^0,\pi^{3})-J(s^0,\pi^0)   >\alpha\gamma_m^{t_{\max}}-\alpha\gamma_m^{t_{\max}}(1-\gamma_m)>0.
		\end{aligned}
	\end{equation}
	Therefore, for all $\pi^{3}\in\Pi^{3}$ and $\pi^{0}\in\Pi^{0}$, we can conclude that $J(s^0,\pi^{3})>J(s^0,\pi^0)$.
	
	$\mathbf{Step\ 5.\ Prove\ that}$ $J(s^0,\pi^{3})>J(s^0,\pi^{1})$, $\mathbf{for\ all}$ $\pi^{3}\in\Pi^{3}$ $\mathbf{and}$ $\pi^{1}\in\Pi^{1}.$ Considering $J(s^0,\pi^{1})$, based on the definition of policy set $\Pi^1$ and rewards (\ref{eq:arreq}) and (\ref{eq:safeeq}), for all $\pi^{1}\in\Pi^{1}$, we have
	\begin{equation}
		\left\{\begin{array}{ll}
			&J_m^p(s^0,\pi^{1})=0,\\
			&J_m^{a}(s^0,\pi^{1})=0.
		\end{array}\right.\label{eq:JN1}
	\end{equation}
	Replace (\ref{eq:JN1}) into (\ref{eq:Jall}), we obtain
	\begin{align}
		J(s^0,\pi^{1}) &=J^{g}_m(s^0,\pi^{1}). \label{eq:pi1}
	\end{align}
	In accordance with (\ref{eq:pi1}) and (\ref{eq:pi3}), it can be obtained
	\begin{equation}
		J(s^0,\pi^{3})-J(s^0,\pi^{1})  >\alpha\gamma_m^{t_{\max}}+J_m^{g}(s^0,\pi^{3})-J_m^{g}(s^0,\pi^{1}).
	\end{equation}
	According to equation (\ref{eqjnupper}), we have 
	\begin{equation}
		J(s^0,\pi^{3})-J(s^0,\pi^{1})   >\alpha\gamma_m^{t_{\max}}-\alpha\gamma_m^{t_{\max}}(1-\gamma_m)>0.
	\end{equation}
	Here, we can infer that $J(s^0,\pi^{3})>J(s^0,\pi^{1})$, for all $\pi^{3}\in\Pi^{3}$ and $\pi^{1}\in\Pi^{1}$.
	
	$\mathbf{Step\ 6.\ Categorize\ the\ analysis\ based\ on\ the\ final}$  $\mathbf{time\ under\ policy\ set}$ $\Pi^2$ $\mathbf{and}$ $\Pi^3$. Without loss of generality, let us fix the initial state $s^0$ and consider two cases based on the time required to reach the terminal state following policy sets $\Pi^2$ and $\Pi^3$:
	\begin{itemize}[itemsep=0pt,topsep=0pt,parsep=0pt, leftmargin=*]
		\item Case 1. $T_{\pi^2_1}\geq T_{\pi^3_1}$, where $T_{\pi^2_1}$ denotes the time required to satisfy\ the \ terminal\ constraint under policy $\pi^2_1\in \Pi^2$, and $T_{\pi^3_1}$ denotes the time required under policy $\pi^3_1\in\Pi^{3}$.
		\item Case 2. $T_{\pi^2_2}< T_{\pi^3_2}$, where $T_{\pi^2_2}$ denotes the time required to satisfy\ the \ terminal\ constraint under policy $\pi^2_2\in \Pi^2$, and $T_{\pi^3_2}$ denotes the time required under policy $\pi^3_2\in\Pi^{3}$.
	\end{itemize}
	Below, for the above two cases, we will sequentially demonstrate that for any $\pi^2\in\Pi^{2}$ and any $\pi^3\in\Pi^{3}$, we have $J(s^0,\pi^{2}) < J(s^0,\pi^{3})$.
	
	$\mathbf{Step\ 7.\ Prove\ that}$ $J(s^0,\pi^3_1) >J(s^0,\pi^2_1)$, $\mathbf{for\ all}$ $\pi^3_1\in\Pi^{3}$ $\mathbf{and}$ $\pi^2_1\in \Pi^{2}$ $\mathbf{with}$ $T_{\pi^2_1}\geq T_{\pi^3_1}$.
	For any $\pi^2_1\in \Pi^{2}$ and any $\pi^3_1\in\Pi^{3}$ satisfying $T_{\pi^2_1}\geq T_{\pi^3_1}$, we have
	\begin{equation}
		\left\{\begin{array}{ll}
			&J_m^a(s^0,\pi^2_1)=\alpha\gamma_m^{T_{\pi^2_1}},\\
			&J_m^a(s^0,\pi^3_1)=\alpha\gamma_m^{T_{\pi^3_1}}. 
		\end{array}\right.
		\label{eqpi23}
	\end{equation}
	Since $T_{\pi^2_1}\geq T_{\pi^3_1}$, the inequality $\alpha\gamma_m^{T_{\pi^2_1}} \leq \alpha\gamma_m^{T_{\pi^3_1}}$ holds. Substituting this inequality into (\ref{eqpi23}), we obtain
	\begin{align}
		J_m^a(s^0,\pi^2_1) - J_m^a(s^0,\pi^3_1)\leq 0. \label{eqja2131}
	\end{align}
	Based on the characteristics of $\Pi^2$ and $\Pi^3$, we observe that the state constraints cannot be satisfied following policy set $\Pi^2_1$, whereas they can be met with policy set $\Pi^3_1$. Therefore, we have
	\begin{equation}
		\left\{\begin{array}{ll}
			&J_m^p(s^0,\pi^2_1)<-\lambda\gamma_m^{t_{\max}} ,\\
			&J_m^p(s^0,\pi^3_1)=0. 
		\end{array}\right.
		\label{eqpic}
	\end{equation}
	According to (\ref{eqpic}), we obtain
	\begin{equation}
		J_m^p(s^0,\pi^2_1)-J_m^p(s^0,\pi^3_1)<-\gamma_m^{t_{\max}}\lambda. \label{j23cons}
	\end{equation}
	Based upon (\ref{eqjnupper}), we have
	\begin{equation}
		J_m^g(s^0,\pi^2_1)- J_m^g(s^0,\pi^3_1)<\alpha\gamma_m^{t_{\max}}(1-\gamma_m)<\alpha\gamma_m^{t_{\max}}.\label{j23n}
	\end{equation}
	Below, we calculate $J(s^0,\pi^2_1)- J(s^0,\pi^3_1)$. Referring to (\ref{eq:Jall}), we have
	\begin{equation}
		\begin{aligned}
			J(s^0,\pi^2_1)- J(s^0,\pi^3_1) =&J_m^a(s^0,\pi^2_1)-J_m^a(s^0,\pi^3_1)\\
			&+J_m^g(s^0,\pi^2_1)-J_m^g(s^0,\pi^3_1)\\
			&+J_m^p(s^0,\pi^2_1)-J_m^p(s^0,\pi^3_1).
		\end{aligned}
		\label{eq2131}
	\end{equation}
	Incorporating (\ref{eqja2131}),  (\ref{j23cons}) and (\ref{j23n}) into (\ref{eq2131}), we obtain
	\begin{equation}
		J(s^0,\pi^2_1)- J(s^0,\pi^3_1)<\gamma_m^{t_{\max}}(\alpha-\lambda).\label{eqp10}
	\end{equation}
	Given that the inequality $\lambda>\alpha\gamma_m^{t_{c}-t_{\max}}$ in (\ref{eqpenbeta1}) holds, we can substitute it into (\ref{eqp10}) to obtain
	\begin{equation}
		\begin{aligned}
			J(s^0,\pi^2_1)- J(s^0,\pi^3_1) 
			<&\gamma_m^{t_{\max}}(\alpha-\alpha\gamma_m^{t_{c}-t_{\max}})\\
			=&\gamma_m^{t_{\max}}\alpha(1-\gamma_m^{t_{c}-t_{\max}}).
		\end{aligned}
	\end{equation}
	In that $t_{c}-t_{\max}\leq 0$ and $0<\gamma_m<1$, we have $\gamma_m^{t_{c}-t_{\max}}\geq1$. Thus, it can be concluded that
	\begin{align}
		J(s^0,\pi^2_1)- J(s^0,\pi^3_1) 
		<&0. \label{eqj2131end}
	\end{align}
	Hence, we can assert that $J(s^0,\pi^3_1) >J(s^0,\pi^2_1)$, for all $\pi^{3}_1\in\Pi^{3}$ and $\pi^{2}_1\in\Pi^{2}$ with $T_{\pi^{2}_1}\geq T_{\pi^{3}_1}$.
	
	$\mathbf{Step\ 8.\ Prove\ that}$ $J(s^0,\pi^{3}_2) >J(s^0,\pi^{2}_2)$, $\mathbf{for\ all}$ $\pi^{3}_2\in\Pi^{3}$ $\mathbf{and}$ $\pi^{2}_2\in\Pi^{2}$ $\mathbf{with}$ $T_{\pi^{2}_2}< T_{\pi^{3}_2}$. For any $\pi^{2}_2\in\Pi^{2}$ and any $\pi^{3}_2\in\Pi^{3}$ with $T_{\pi^{2}_2}< T_{\pi^{3}_2}$, the derivation of (\ref{j23cons}) and (\ref{j23n}) remains valid for policies $\Pi_{2}^2$ and $\Pi_{3}^2$. Thus, we have
	\begin{equation}
		\left\{\begin{array}{ll}
			&J_m^p(s^0,\pi^{2}_2)-J_m^p(s^0,\pi^{3}_2)<-\gamma_m^{t_{\max}}\lambda ,\\
			&J_m^g(s^0,\pi^{2}_2)- J_m^g(s^0,\pi^{3}_2)<\gamma_m^{t_{\max}}\alpha. 
		\end{array}\right.
		\label{eqconsn2232}
	\end{equation}
	Next, let's analyze the magnitude of $J_m^a(s^0,\pi^{2}_2) - J_m^a(s^0,\pi^{3}_2)$. Firstly, let's examine the magnitude of $J_m^a(s^0,\pi^{2}_2)$. According to the definition of $t_{c}$ in (\ref{eqtc}), we have $T_{\pi^{2}_2}\geq t_{c} $. Hence, we obtain
	\begin{equation}
		\begin{aligned}
			J_m^a(s^0,\pi^{2}_2)=\gamma_m^{T_{\pi^{2}_2}}\alpha
			\leq\gamma_m^{t_{c}}\alpha.
		\end{aligned}
		\label{eqja22}
	\end{equation}
	Let's assess the magnitude of $J_m^a(s^0,\pi^{3}_2)$. Based on the definition of the terminal step $T_f$, we have $T_{\pi^{3}_2} \leq t_{\max}$. Therefore, we can infer that
	\begin{equation}
		\begin{aligned}
			J_m^a(s^0,\pi^{3}_2)=\gamma_m^{T_{\pi^{3}_2}}\alpha
			\geq\gamma_m^{t_{\max}}\alpha.\label{eqja32}
		\end{aligned}
	\end{equation}
	Referring to (\ref{eqja22}) and (\ref{eqja32}), we have
	\begin{align}
		J_m^a(s^0,\pi^{2}_2)-J_m^a(s^0,\pi^{3}_2)\leq&\gamma_m^{t_{c}}\alpha-\gamma_m^{t_{\max}}\alpha.\label{eqja2232}
	\end{align}
	According to (\ref{eq:Jall}), we have
	\begin{equation}
		\begin{aligned}
			J(s^0,\pi^{2}_2)- J(s^0,\pi^{3}_2) =&J_m^a(s^0,\pi^{2}_2)-J_m^a(s^0,\pi^{3}_2)\\
			&+J_m^g(s^0,\pi^{2}_2)-J_m^g(s^0,\pi^{3}_2)\\
			&+J_m^p(s^0,\pi^{2}_2)-J_m^p(s^0,\pi^{3}_2)
		\end{aligned}
		\label{eq2232}
	\end{equation}
	Substituting (\ref{eqconsn2232}) and (\ref{eqja2232}) into (\ref{eq2232}), it holds that
	\begin{equation}
		\begin{aligned}
			J(s^0,\pi^{2}_2)- J(s^0,\pi^{3}_2) <&-\gamma_m^{t_{\max}}\lambda+\gamma_m^{t_{c}}\alpha
		\end{aligned}
		\label{eqj2232}
	\end{equation}
	Substituting the inequality $\lambda>\alpha\gamma_m^{t_{c}-t_{\max}}$ in (\ref{eqpenbeta1}) into (\ref{eqj2232}), we obtain
	\begin{equation}
		J(s^0,\pi^{2}_2)- J(s^0,\pi^{3}_2)<0 \label{eq2232end}
	\end{equation}
	Combining equations (\ref{eqj2131end}) and (\ref{eq2232end}), we can conclude that $J(s^0,\pi^{3}) >J(s^0,\pi^{2})$, for all $\pi^{3}\in\Pi^{3}$ and $\pi^{2}\in\Pi^{2}$.
	
	Thus, it can be concluded that for all $\pi^3\in\Pi^3$, and for all $\pi^{0,1,2}\in \Pi^0\cup\Pi^1\cup \Pi^2$, it holds that $J(s^0,\pi^{3})>J(s^0,\pi^{0,1,2})$. Next, we will proceed to prove that for all $\pi^3\in\Pi^3$ and $\overline\pi^3\in\Pi^3$, if $T_{\pi^3}<T_{\overline\pi^3}$ then $J(s^0,\pi^{3})> J(s^0,\overline\pi^3)$. 
	
	$\mathbf{Step\ 9.\ Prove\ that}$ $J(s^0,\pi^{3}) >J(s^0,\overline\pi^3)$, $\mathbf{for\ all}$ $\pi^{3}\in\Pi^{3}$ $\mathbf{and}$ $\overline\pi^{3}\in\Pi^{3}$ $\mathbf{with\ terminal\ time}$ $T_{\overline\pi^3}>T_{\pi^{3}}$.
	We analyze the components of $J(s^0,\pi^{3})$ and $J(s^0,\overline\pi^3)$ sequentially. Since the agent takes time $T_{\pi^{3}}$ to reach the terminal state following policy $\pi^{3}\in\Pi_3$, and time $T_{\overline\pi^3}$ following policy $\overline\pi^3\in\Pi^{3}$, we have
	\begin{equation}
		\left\{\begin{array}{ll}
			&J_m^a(s^0,\pi^{3})=\alpha\gamma_m^{T_{\pi^{3}}},\\
			&J_m^a(s^0,\overline\pi^{3})=\alpha\gamma_m^{T_{\overline\pi^3}}. 
		\end{array}\right.
		\label{eqja3l3}
	\end{equation}
	Considering the properties of policy set $\Pi_3$, the state constraints can be satisfied following policies $\pi^{3}$ and $\overline\pi^{3}$. Therefore, we have
	\begin{equation}
		\left\{\begin{array}{ll}
			&J_m^p(s^0,\pi^{3})=0 ,\\
			&J_m^p(s^0,\overline\pi^{3})=0. 
		\end{array}\right.
		\label{eqjl3l3}
	\end{equation}
	According to (\ref{eqjnupper}), we have
	\begin{equation}
		J_m^g(s^0,\pi^{3})- J_m^g(s^0,\overline\pi^{3})>-\alpha\gamma_m^{t_{\max}}(1-\gamma_m)
		\label{eqN3L3}
	\end{equation}
	Based on (\ref{eq:Jall}), we have
	\begin{equation}
		\begin{aligned}
			J(s^0,\pi^{3})- J(s^0,\overline\pi^3) =&J_m^a(s^0,\pi^{3})-J_m^a(s^0,\overline\pi^{3})\\
			&+J_m^g(s^0,\pi^{3})-J_m^g(s^0,\overline\pi^{3})\\
			&+J_m^p(s^0,\pi^{3})-J_m^p(s^0,\overline\pi^{3})
		\end{aligned}
		\label{eq3L3}
	\end{equation}	
	Substituting equation (\ref{eqja3l3}),  (\ref{eqjl3l3}) and (\ref{eqN3L3}) into (\ref{eq3L3}), we obtain	
	\begin{equation}
		J(s^0,\pi^{3})- J(s^0,\overline\pi^3)>\alpha(\gamma_m^{T_{\pi^{3}}}-\gamma_m^{T_{\overline\pi^3}})-\alpha\gamma_m^{t_{\max}}(1-\gamma_m)
		\label{eqp313}
	\end{equation}
	Given that $T_{\pi^{3}}<T_{\overline\pi^3}$, we can confirm that $ T_{\overline\pi^3}-T_{\pi^{3}}\geq1$. Substituting this inequality into (\ref{eq3L3}), we obtain
	\begin{equation}
		\begin{aligned}
			&J(s^0,\pi^{3})- J(s^0,\overline\pi^3)\\
			>&\alpha\gamma_m^{T_{\pi^{3}}}(1-\gamma_m^{T_{\overline\pi^3}-T_{\pi^{3}}})-\alpha\gamma_m^{t_{\max}}(1-\gamma_m) \\
			\geq&\alpha\gamma_m^{t_{\max}}(1-\gamma_m)-\alpha\gamma_m^{t_{\max}}(1-\gamma_m)\\
			=&0
		\end{aligned}
	\end{equation}
	
	Thus, we can conclude that for all $\pi^3\in\Pi^3$ and $\overline\pi^3\in\Pi^3$, if $T_{\pi^{3}}<T_{\overline\pi^3}$ then $J(s^0,\pi^{3})> J(s^0,\overline\pi^3)$. In conclusion, it can be summarized that among non-constraints-violating policies $\pi^{3}\in\Pi^{3}$, the policy $\pi^{3*}\in\Pi^{3}$ with the minimal terminal time has the maximum return. Thus, the optimal policy obtained based on (\ref{eq:rewardt}) satisfies the \textit{minimum-time problem}.
	
	\section{Proof of Corollary 1}\label{sec:C1}
	Corollary 1 addresses the \textit{minimum-time problem} without state constraints, which is a special case of the \textit{minimum-time problem} where the feasible state set \( C = \mathbb{R}^{np} \) (\ref{safe2}). This connection allows us to apply properties from Theorem 2 for the \textit{minimum-time problem} to Corollary 1 for the unconstrained case. 
	
	We first categorize joint policies into two main sets, \(\Pi^i, i = 4,5\), as shown in Table \ref{table12}. We then establish the relationship between policy sets \(\Pi^4\) and \(\Pi^1\), and \(\Pi^5\) and \(\Pi^3\), where policy sets \(\Pi^3\) and \(\Pi^1\) are the ones considered in Theorem 2 of Table \ref{table1}. Given the feasible state set \(C = \mathbb{R}^{np}\) in this corollary, the state constraint (\ref{safe2}) is always satisfied under \(\Pi^i, i = 4,5\). According to the definition of policy set \(\Pi^i, i = 1,3,4,5\) in Tables \ref{table12} and \ref{table1}, we find that \(\Pi^4\) and \(\Pi^1\), and \(\Pi^5\) and \(\Pi^3\), are consistent in satisfying state and terminal constraints. The only difference between the POMDPs in Corollary 1 and Theorem 2 is the penalty weight \(\lambda\) as shown in (\ref{eqpenbeta2}) and (\ref{eqpenbeta1}), and because the state constraint is satisfied under \(\Pi^i, i = 1,3,4,5\), the differing \(\lambda\) values do not impact the return under these policy sets. Thus, the return ranges of \(\Pi^1\) and \(\Pi^4\), which both satisfy the terminal constraint, are consistent, as are those of \(\Pi^3\) and \(\Pi^5\), which do not.
	
	\begin{table}[htbp]
		\centering  
		\vspace{-5pt} 
		\caption{Policy classification}  
		\fontsize{7.5pt}{6.5pt}\selectfont 
		\label{table12}  
		\begin{tabular}{|c|c|}  
			\hline 
			&Satisfy terminal constraint\\  
			\hline
			$\Pi^4$&$\times$ \\
			\hline
			$\Pi^5$&$\checkmark$ \\
			\hline
		\end{tabular}
		\vspace{-5pt} 
	\end{table}
	
	Next, building on Theorem 2, we demonstrate the relationship between the returns of policies in \(\Pi^4\) and \(\Pi^5\). From \textbf{Step 5} of Theorem 2, we have \(J(s^0, \pi^3) > J(s^0, \pi^1)\) for all \(\pi^3 \in \Pi^3\) and \(\pi^1 \in \Pi^1\). Given the consistency between \(\Pi^1\) and \(\Pi^4\), and \(\Pi^3\) and \(\Pi^5\), it follows that \(J(s^0, \pi^5) > J(s^0, \pi^4)\) for all \(\pi^5 \in \Pi^5\) and \(\pi^4 \in \Pi^4\). According to \textbf{Step 9} of Theorem 2, \(J(s^0, \pi^3) > J(s^0, \overline{\pi}^3)\) for all \(\pi^3 \in \Pi^3\) and \(\overline{\pi}^3 \in \Pi^3\) where \(T_{\overline{\pi}^3} > T_{\pi^3}\). By the consistency of \(\Pi^3\) and \(\Pi^5\), we have \(J(s^0, \pi^5) > J(s^0, \overline{\pi}^5)\) for all \(\pi^5 \in \Pi^5\) and \(\overline{\pi}^5 \in \Pi^5\) with \(T_{\overline{\pi}^5} > T_{\pi^5}\). In conclusion, among policies \(\pi^5 \in \Pi^5\) that do not violate terminal constraints, the policy \(\pi^{5*} \in \Pi^5\) with the minimal terminal time maximizes the return. Thus, the optimal policy derived from (\ref{eq:rewardt}) meets the \textit{minimum-time problem without state constraints}.
	
\end{document}